\documentclass[journal]{IEEEtran}

\usepackage{lipsum} 

\usepackage{comment}

\usepackage{cite}

\ifCLASSINFOpdf
  \usepackage[pdftex]{graphicx}
\else
\fi

\usepackage{color}

\usepackage{amsmath,amssymb,amsthm,mathtools}
\usepackage{bm}
\usepackage{bbm}
\usepackage{mathrsfs}
\usepackage{upgreek}
\usepackage{soul}
\usepackage{setspace}
\usepackage{float}
\usepackage{colortbl}
\usepackage{algorithm, algpseudocode}

\usepackage{slashbox}

\ifCLASSOPTIONcompsoc
  \usepackage[caption=false,font=normalsize,labelfont=sf,textfont=sf]{subfig}
\else
  \usepackage[caption=false,font=footnotesize]{subfig}
\fi

\usepackage[acronym,shortcuts]{glossaries}

\newacronym{3GPP}{3GPP}{3rd generation partnership project}
\newacronym{5G}{5G}{fifth-generation}
\newacronym{5G NR}{5G NR}{fifth-generation new radio}
\newacronym{6G}{6G}{sixth-generation}
\newacronym{AoA}{AoA}{angle of arrival}
\newacronym{AoD}{AoD}{angle of departure}
\newacronym{ADC}{ADC}{analog-to-digital converter}
\newacronym{ADR}{ADR}{antenna decentralized ratio}
\newacronym{AER}{AER}{activity error rate}
\newacronym{AMP}{AMP}{approximate message passing}
\newacronym{ANN}{ANN}{approximate nearest neighbor}
\newacronym{AR}{AR}{auto-regressive}
\newacronym{AP}{AP}{access point}
\newacronym{AUD}{AUD}{active user detection}
\newacronym{AWGN}{AWGN}{additive white Gaussian noise}
\newacronym{BAd-VAMP}{BAd-VAMP}{bilinear adaptive VAMP}
\newacronym{BMMSE}{BMMSE}{Bussgang minimum mean square error}
\newacronym{BER}{BER}{bit error rate}
\newacronym{BiGAMP}{BiGAMP}{bilinear generalized approximate message passing}
\newacronym{BiGaBP}{BiGaBP}{bilinear Gaussian belief propagation}
\newacronym{BLER}{BLER}{block error rate}
\newacronym{BP}{BP}{belief propagation}
\newacronym{BS}{BS}{base station}
\newacronym{CAP}{CAP}{central AP}
\newacronym{CCU}{CCU}{central computing unit}
\newacronym{CDF}{CDF}{cumulative distribution function}
\newacronym{CDL}{CDL}{clustered delay line}
\newacronym{CLT}{CLT}{central limit theorem}
\newacronym{CPU}{CPU}{central processing unit}
\newacronym{CE}{CE}{channel estimation}
\newacronym{CP}{CP}{channel prediction}
\newacronym{CT}{CT}{channel tracking}
\newacronym{CP-JCDE}{CP-JCDE}{CP-aided JCDE}
\newacronym{CF-mMIMO}{CF-mMIMO}{cell-free massive MIMO}
\newacronym{CSI}{CSI}{channel state information}
\newacronym{CSIDCO}{CSIDCO}{complex SIDCO}
\newacronym{DCC}{DCC}{dynamic cooperation clustering}
\newacronym{DD}{DD}{data detection}
\newacronym{DFT}{DFT}{discrete Fourier transform}
\newacronym[longplural={degrees of freedom}]{DoF}{DoF}{degrees of freedom}
\newacronym{DQ}{DQ}{de-quantization}
\newacronym{DNN}{DNN}{deep neural network}
\newacronym{DQN}{DQN}{deep Q-network}
\newacronym{DQL}{DQL}{deep Q learning}
\newacronym{DRL}{DRL}{deep reinforcement learning}
\newacronym{DU}{DU}{deep unfolding}
\newacronym{eMBB}{eMBB}{enhanced mobile broadband}
\newacronym{ECF}{ECF}{estimate-compress-forward}
\newacronym{EP}{EP}{expectation propagation}
\newacronym{EXIT}{EXIT}{extrinsic information transfer}
\newacronym{FA}{FA}{false alarm}
\newacronym{FFT}{FFT}{fast Fourier transform}
\newacronym{FN}{FN}{factor node}
\newacronym{FG}{FG}{factor graph}
\newacronym{GaBP}{GaBP}{Gaussian belief propagation}
\newacronym{GAMP}{GAMP}{generalized approximate message passing}
\newacronym{GF}{GF}{grant-free}
\newacronym{GM}{GM}{Gaussian-mixture}
\newacronym{GPU}{GPU}{graphics processing unit}
\newacronym{IDD}{IDD}{iterative detection and decoding}
\newacronym{IFFT}{IFFT}{inverse fast Fourier transform}
\newacronym{IVF}{IVF}{inverted file}
\newacronym{i.i.d.}{i.i.d.}{independent and identically distributed}
\newacronym{IoT}{IoT}{internet of things}
\newacronym{JACDE}{JACDE}{joint activity, channel and data estimation}
\newacronym{JCTDD}{JCTDD}{joint channel tracking and data detection}
\newacronym{JACE}{JACE}{joint activity and channel estimation}
\newacronym{JCDE}{JCDE}{joint channel and data estimation}
\newacronym{LDPC}{LDPC}{low-density parity-check}
\newacronym{LLR}{LLR}{log-likelihood ratio}
\newacronym{LKF}{LKF}{linear Kalman filter}
\newacronym{LMMSE}{LMMSE}{linear minimum mean-square error}
\newacronym{LSA}{LSA}{latent semantic analysis}
\newacronym{LS}{LS}{least square}
\newacronym{KF}{KF}{Kalman filter}
\newacronym{KL}{KL}{Kullback-Leibler}
\newacronym{MAC}{MAC}{multiple-access channel}
\newacronym{MPDQ}{MPDQ}{message passing DQ}
\newacronym{MCS}{MCS}{modulation and coding scheme}
\newacronym{MD}{MD}{miss-detection}
\newacronym{MF}{MF}{matched filter}
\newacronym{MFB}{MFB}{Matched filter bound}
\newacronym{MNS}{MNS}{minimum norm solution}
\newacronym{MI}{MI}{mutual information}
\newacronym{MIMO}{MIMO}{multiple-input multiple-output}
\newacronym{MIMO-OFDM}{MIMO-OFDM}{multiple-input multiple-output orthogonal frequency division multiplexing}
\newacronym{MU-MIMO}{MU-MIMO}{multi-user multiple-input multiple-output}
\newacronym{MU-MIMO-OFDM}{MU-MIMO-OFDM}{multi-user multiple-input multiple-output orthogonal frequency-division multiplexing}
\newacronym{mMIMO}{mMIMO}{massive multiple-input multiple-output}
\newacronym{mmWave}{mmWave}{millimeter-wave}
\newacronym{ML}{ML}{machine learning}
\newacronym{MMSE}{MMSE}{minimum mean-square error}
\newacronym{mMTC}{mMTC}{massive machine type communications}
\newacronym{MMV-AMP}{MMV-AMP}{multiple measurement vector approximate message passing}
\newacronym{MSE}{MSE}{mean square error}
\newacronym{MUD}{MUD}{multi-user detection}
\newacronym{NR}{NR}{new radio}
\newacronym{NMSE}{NMSE}{normalized mean square error}
\newacronym{O-RAN}{O-RAN}{open radio access network}
\newacronym{OLLA}{OLLA}{outer-loop link adaptation}
\newacronym{OFDM}{OFDM}{orthogonal frequency division multiplexing}
\newacronym{PDA}{PDA}{probabilistic data association}
\newacronym{PDF}{PDF}{probability density function}
\newacronym{PE}{PE}{prediction error}
\newacronym{PMF}{PMF}{probability mass function}
\newacronym{PPP}{PPP}{Poisson point process}
\newacronym{PQ}{PQ}{product quantization}
\newacronym{QAM}{QAM}{quadrature amplitude modulation}
\newacronym{QP}{QP}{quadratic program}
\newacronym{QPSK}{QPSK}{quadrature phase-shift keying}
\newacronym{RB}{RB}{resource block}
\newacronym{RL}{RL}{reinforcement learning}
\newacronym{SIDCO}{SIDCO}{sequential iterative decorrelation via convex optimization}
\newacronym{SD}{SD}{sphere decoding}
\newacronym{SGA}{SGA}{scalar Gaussian approximation}
\newacronym{SIC}{SIC}{soft interference cancellation}
\newacronym{SINR}{SINR}{signal-to-interference-plus-noise ratio}
\newacronym{SIR}{SIR}{signal-to-interference ratio}
\newacronym{SNR}{SNR}{signal-to-noise ratio}
\newacronym{soft IC}{soft IC}{soft interference cancellation}
\newacronym{SotA}{SotA}{state-of-the-art}
\newacronym{SPA}{SPA}{sum-product algorithm}
\newacronym{SVD}{SVD}{singular value decomposition}
\newacronym{SBL}{SBL}{sparse Bayesian learning}
\newacronym{BBI}{BBI}{Bayesian bilinear inference}
\newacronym{TB}{TB}{transport block}
\newacronym{TX}{TX}{transmit}
\newacronym{RX}{RX}{receive}
\newacronym{UE}{UE}{user equipment}
\newacronym{UPA}{UPA}{uniformly spaced planar antenna array}
\newacronym{URA}{URA}{uniform rectangular array}
\newacronym{URLLC}{URLLC}{ultra-reliable low-latency communications}
\newacronym{V2X}{V2X}{vehicle-to-everything}
\newacronym{V2V}{V2V}{vehicle-to-vehicle}
\newacronym{V2I}{V2I}{vehicle-to-infrastructure}
\newacronym{VRU}{VRU}{vulnerable road user}
\newacronym{VAMP}{VAMP}{vector AMP}
\newacronym{VDB}{VDB}{vector database}
\newacronym{VGA}{VGA}{vector Gaussian approximation}
\newacronym{VN}{VN}{variable node}
\newacronym{VSS}{VSS}{vector similarity search}
\newacronym{XL-MIMO}{XL-MIMO}{extra-large MIMO}
\newacronym{ZF}{ZF}{zero-forcing}
\newacronym{flops}{flops}{floating point operations}
\newacronym{ILLA}{ILLA}{inner-loop link adaptation}

\theoremstyle{definition}

\begin{document}
%

\title{Vector Similarity Search-Based MCS Selection\\ in Massive Multi-User MIMO-OFDM}

\author{
Fuga~Kobayashi,~\IEEEmembership{Graduate Student Member,~IEEE,}
Takumi~Takahashi,~\IEEEmembership{Member,~IEEE,}

Shinsuke~Ibi,~\IEEEmembership{Senior Member,~IEEE,}
Takanobu~Doi,~\IEEEmembership{Member,~IEEE,}
Kazushi~Muraoka,~\IEEEmembership{Member,~IEEE,}
and~Hideki~Ochiai,~\IEEEmembership{Fellow,~IEEE}
\thanks{
This work was supported in part by JSPS KAKENHI under Grant JP23K13335, JP23K22754, and JP25H01111; in part by JST, CRONOS, Japan under Grant JPMJCS24N1; and in part by MIC/FORWARD under Grant JPMI240710001. \textit{(Corresponding author: Takumi Takahashi.)}


F. Kobayashi, T. Takahashi, and H. Ochiai are with Graduate School of Engineering, The University of Osaka 2-1 Yamada-oka, Suita, 565--0871, Japan (e-mail: kobayashi-f@wcs.comm.eng.osaka-u.ac.jp, \{takahashi, ochiai\}@comm.eng.osaka-u.ac.jp).

S. Ibi is with Faculty of Science and Engineering, Doshisha University 1-3 Tataramiyakodani, Kyotanabe, 610--0394, Japan (e-mail: sibi@mail.doshisha.ac.jp).

T. Doi and K. Muraoka are with NEC Corporation, 1753 Shimonumabe, Nakahara-ku, Kawasaki, Kanagawa 211--8666, Japan (e-mail: \{doi-takanobu, k-muraoka\}@nec.com).
}
\vspace{-7mm}
}

\markboth{Journal of \LaTeX\ Class Files,~Vol.~14, No.~8, August~2021}%
{Shell \MakeLowercase{\textit{et al.}}: A Sample Article Using IEEEtran.cls for IEEE Journals}
%



\IEEEtitleabstractindextext{%
\begin{abstract}
This paper proposes a novel \ac{MCS} selection framework that integrates \ac{MI} prediction based on \ac{VSS} for massive \ac{MU-MIMO-OFDM} systems with advanced uplink \ac{MUD}.
The framework performs \ac{MCS} selection at the \ac{TB}-level \ac{MI} and establishes the mapping from post-\ac{MUD} \ac{MI} to post-decoding \ac{BLER} using a prediction function generated from \ac{EXIT} curves.
A key innovation is the \ac{VSS}-based \ac{MI} prediction scheme, which addresses the challenge of analytically predicting \ac{MI} in iterative detectors such as \ac{EP}.
In this scheme, an offline \ac{VDB} stores feature vectors derived from \ac{CSI} and average received \ac{SNR}, together with corresponding \ac{MI} values achieved with advanced \ac{MUD}.
During online operation, an \ac{ANN} search on \acp{GPU} enables ultra-fast and accurate \ac{MI} prediction, effectively capturing iterative detection gains.
Simulation results under \ac{5G NR}-compliant settings demonstrate that the proposed framework significantly improves both system and user throughput, ensuring that the detection gains of advanced \ac{MUD} are faithfully translated into tangible system-level performance improvements.

\end{abstract}
\vspace{-2mm}
}

\maketitle

\IEEEdisplaynontitleabstractindextext

%
\IEEEpeerreviewmaketitle

\glsresetall


\section{Introduction}
\label{sec:intro}

Massive \ac{MU-MIMO-OFDM} is a key technology to cope with the explosive growth in the number of uplink \ac{UE} devices in future wireless networks~\cite{Tataria2021,Jiang2021,Chen2023}.
Leveraging the large antenna array at the \ac{BS} enables multiplexing of uplink traffic from multiple UEs in both spatial and frequency domains, significantly improving spectral efficiency and increasing the number of supported \acp{UE}~\cite{Hussein2024,Ito2024TWC,Ito2025JSTSP}.
To separate these spatially multiplexed signals, the \ac{BS} requires \ac{MUD} with low complexity and high accuracy~\cite{Chockalingam2014,Yang2015MIMO}.
Spatial filtering methods such as \ac{ZF} and \ac{LMMSE} detection are widely used for this purpose in \acp{MUD}, but their performance degrades as the ratio of transmit streams to the receive antennas increases.
Advanced \ac{MUD} algorithms, offering a better compromise between complexity and performance, have therefore attracted increasing attention~\cite{Cespedes2014,Takahashi2019_tcom,Furudoi2024ArXiv}.

Maximizing the uplink throughput of \ac{MU-MIMO-OFDM} systems requires adaptive assignment of the \ac{MCS} index to each \ac{UE} based on wireless channel quality~\cite{Martín-Vega2021,Qing2023}.
A common channel quality metric is the predicted subcarrier-wise \ac{SINR} achievable with \ac{MUD}~\cite{Shikida2016,Doi2023,Srinath2023}.
The \textit{effective} \ac{SNR} has also been widely adopted for \ac{MCS} selection~\cite{Cipriano2008,Francis2014}.
%
%
However, in many standardization specifications~\cite{Wifi_2016,LTE_Coding,5GNR_Coding}, channel coding is applied per \ac{TB}, which spans multiple time slots and subcarriers, resulting in a single codeword distributed over time-frequency resources with varying quality.
%
%
Therefore, it is essential to evaluate \ac{TB}-level channel quality while accounting for frequency-selective variations.
\Ac{MI}, which can be computed per \ac{TB}, has been recognized as a more suitable metric for this purpose~\cite{Sayana2007,Jensen2010,Hu2023}.

With spatial filtering methods such as \ac{LMMSE} detection, the post-\ac{MUD} \ac{SINR} can be analytically estimated from the \ac{CSI}.
The estimated subcarrier-wise \ac{SINR} is then converted to \ac{MI}, and the predicted \ac{MI} of the demodulator output is obtained by averaging over all subcarriers within the allocated resource (\textit{i.e.}, one \ac{TB}).
In contrast, with nonlinear iterative detection such as \ac{EP}~\cite{Minka2013,Rangan2019,Takeuchi2020}, analytical estimation of the post-\ac{MUD} \ac{SINR} is challenging, since the detection accuracy depends on the convergence behavior of the iterative process~\cite{Doi2023,Kobayashi2025}.
%
%
If the \ac{MI} of the demodulator output cannot be predicted accurately, the gains in \ac{MUD} performance from iterative detection cannot be fully utilized in \ac{MCS} selection. 
Consequently, even if advanced iterative \ac{MUD} improves link-level detection accuracy, as reported in the literature~\cite{Wang2020EP,Tamaki2022,Takahashi2022TWC}, it may not translate into higher overall system throughput.

Based on these observations, this paper proposes a novel \ac{VSS}-based offline learning approach for predicting the achievable \ac{MI} of the demodulator output when using advanced \ac{MUD}, employing a \ac{VDB} and \ac{ANN} search~\cite{Cao2018,johnson2019}.
The method consists of two phases: \ac{VDB} construction via offline learning and online \ac{MI} prediction using the constructed \ac{VDB}.
In the offline phase, feature vectors (\textit{keys})—computed from the \ac{CSI} and average received \ac{SNR}—are generated and stored in the \ac{VDB} together with the corresponding \acp{MI} (\textit{values}) obtained from actual transmissions over those channels.
Repeating this process for numerous channel realizations yields a large-scale key-value stored database. 
In the online phase, a feature vector is generated from the estimated \ac{CSI} and average received \ac{SNR}, and \ac{VSS} is performed via \ac{ANN} search on the \ac{VDB} to retrieve the \ac{MI} value associated with the most similar key.
As this value corresponds to the demodulator output \ac{MI} achieved for a similar channel, the prediction inherently reflects the convergence characteristics of iterative \ac{MUD}, enabling high prediction accuracy.
\ac{ANN} search has been employed to perform accurate \ac{VSS} by exploiting similarities or correlations in the data~\cite{pan2024} and has been developed in fields such as cross-language translation and high-dimensional image classification~\cite{Cao2018,johnson2019}.
Recently, high-speed \ac{ANN} libraries capable of executing search algorithms on \acp{GPU}, such as Faiss~\cite{Douze2025}, Milvus~\cite{Wang2021Milvus}, and SONG~\cite{Zhao2020SONG}, have emerged. 
Leveraging these advances, \ac{ANN} search has been applied to wireless communication problems, including transmission power allocation~\cite{Yang2023} and meta-learning for \ac{DU}-aided signal detection~\cite{Wei2024}.
This study can be considered an extension of such applications, targeting \ac{MI} prediction for advanced iterative \ac{MUD}.

The main contributions of this work are as follows:
\begin{itemize}
    \item A novel \ac{VSS}-based \ac{MI} prediction method is developed for \ac{EP}-based iterative detection\footnote{While the proposed framework can be applied to arbitrary MUD schemes, we adopt \ac{EP} as an initial study because of its strong theoretical foundation, excellent practical performance, and stable convergence~\cite{Rangan2019,Takeuchi2020}.}.
    The \ac{VDB} stores feature vectors computed from the \ac{CSI} and average received \ac{SNR}, paired with the measured \ac{MI} obtained via \ac{EP} detection. 
    During prediction, \ac{VSS} retrieves the \ac{MI} that corresponds to the most similar feature vector using \ac{ANN} search. 
    To further enhance prediction accuracy, the \ac{MI} values corresponding to multiple high-similarity vectors can be averaged—referred to as the top-$K$ method. 
    Simulation results demonstrate that the proposed approach achieves high prediction accuracy even with iterative \ac{MUD}, addressing the limitations of conventional \ac{SINR}-based prediction methods.
    \item The decoding characteristics of the channel decoder for each \ac{MCS} index described in the \ac{5G NR} standard are modeled using \ac{EXIT} functions~\cite{ten2001,Hagenauer2004}. 
    The decoding behavior is represented by an \ac{EXIT} curve, which relates the \ac{MI} between the decoder input and output. 
    By replacing the decoder output \ac{MI} with the post-decoding \ac{BLER} in the \ac{EXIT} curve, the post-decoding \ac{BLER} can be predicted from the input \ac{MI} without performing actual decoding. 
    Simulations demonstrate highly accurate prediction of decoding characteristics.
    %
    \item An \ac{MI}-based \ac{MCS} selection framework is established by integrating the above two methods.
    For each modulation scheme, \ac{MI} values are predicted using the \ac{VDB} and \ac{ANN} search based on the estimated \ac{CSI}. 
    The achievable post-decoding \ac{BLER} is then obtained from the predicted \ac{MI} via \ac{EXIT} curves. 
    For each \ac{UE}, the highest \ac{MCS} index that satisfies the target (reference) \ac{BLER} is selected, illustrating the performance gains achieved through \ac{EP}-based advanced \ac{MUD}.
    %
    \item Comprehensive evaluations are conducted, demonstrating significant improvements in both system and user throughput compared with conventional \ac{SINR}-based schemes. 
    The results highlight the importance of selecting \ac{MCS} according to the actual detection performance. 
    An \ac{MCS} selection strategy that incorporates the improved performance of advanced iterative \ac{MUD} yields substantial throughput benefits. 
    Analysis of the selected \ac{MCS} indices under varying channel conditions provides quantitative evidence of the adaptability of the method and its ability to optimize overall system performance.
\end{itemize}

To our knowledge, no previous work has proposed an \ac{MCS} selection framework for massive \ac{MU-MIMO-OFDM} systems that can assess channel quality while accounting for convergence characteristics of advanced iterative \ac{MUD}.
Although numerous studies have improved \ac{MUD} performance in link-level simulations using iterative detection, many such simulations rely on idealized conditions that differ substantially from practical wireless environments.
This work bridges that gap, serving as a crucial first step toward translating theoretical advances in iterative \ac{MUD} into tangible performance gains in real-world wireless communication systems.

\begin{figure*}[t]
\centering
\includegraphics[width=1.8\columnwidth,keepaspectratio=true]{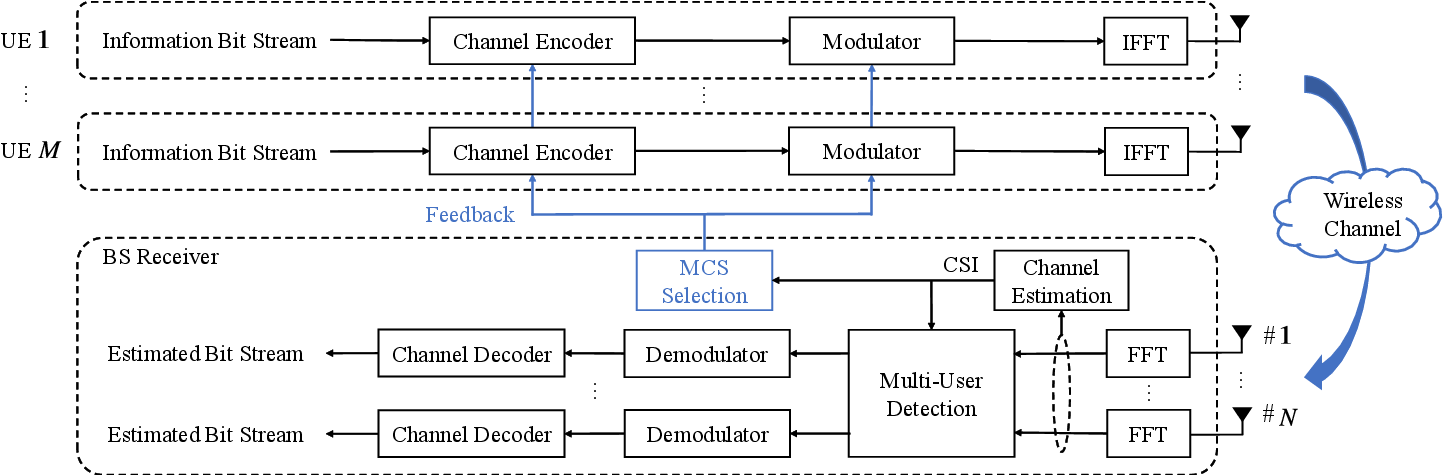}
\caption{Block diagram of the \ac{MU-MIMO-OFDM} system, where the \ac{MCS} selection stage is highlighted in blue.}
\vspace{-3mm}
\label{fig: SystemConfig}
\end{figure*}

The remainder of this paper is organized as follows.
Section~\ref{sec: system model} describes the system model.
Section~\ref{sec: MI-based MCS Selection} presents the \ac{MCS} selection framework using \ac{LMMSE} filtering as \ac{MUD}, where functionalizing decoder characteristics via \ac{EXIT} analysis under the \ac{5G NR} standard constitutes a novel contribution.
It also demonstrates that \ac{LMMSE}-based \ac{MUD} enables analytically accurate \ac{MI} prediction and identifies the challenges posed by \ac{EP}-based iterative \ac{MUD}.
Section~\ref{sec: Proposed Method} addresses these challenges by introducing a \ac{VSS}-based \ac{MI} prediction method that accurately estimates post-\ac{MUD} \ac{MI} for \ac{EP}-based \ac{MUD}.
Section~\ref{sec: Throughput Evaluation} integrates these techniques into the \ac{MCS} selection framework and demonstrates significant improvements in system and user throughput.
Section~\ref{sec: FutherDiscussion} discusses the feasibility and performance implications of the proposed method using dynamic system-level simulations that incorporate long-term variations in channel statistics, integration with \ac{OLLA}, and scheduling mechanisms for resource allocation.
Section~\ref{sec: Conclusion} concludes the paper.

\textit{Notation:} Sets of real and complex numbers are denoted by $\mathbb{R}$ and $\mathbb{C}$, respectively.
Vectors are represented by lower-case boldface letters, and matrices by upper-case boldface letters.
The transpose and conjugate transpose operators are denoted by $\cdot^{\mathsf{T}}$ and $\cdot^{\mathsf{H}}$, respectively.
The $a \times a$ identity matrix is represented by $\bm{I}_a$.
A diagonal matrix with elements of the vector $\bm{a}$ on its main diagonal is denoted by $\mathrm{diag}[\bm{a}]$.
The element in the $i$-th row and $j$-th column of matrix $\bm{A}$ is denoted by $[\bm{A}]_{i,j}$.
A complex Gaussian distribution with mean $a$ and variance $b$ is denoted by $\mathcal{CN}(a, b)$.
Finally, the notation $\mathcal{O}(\cdot)$ denotes the complexity order unless otherwise specified.

\subsection{Related Works}

To the best of our knowledge, there is no prior work that explicitly translates the iterative gain achieved by iterative detectors into throughput improvement through advanced \ac{MCS} selection.
Although theoretical studies focusing on the behavior analysis of detectors themselves, as well as practical studies dedicated to link adaptation, have been reported so far, research that integrates these two aspects remains scarce, despite its fundamental importance.

For example, advanced iterative detectors, including \ac{EP}, have historically evolved alongside asymptotic performance analyzes in the large-system limit, and a vast body of literature exists on their asymptotic optimality~\cite{Rangan2019,Takeuchi2020}.
However, such asymptotic analyzes cannot be directly applied to the statistically non-uniform and practically important finite-dimensional systems considered in this study.
In particular, it is intrinsically difficult to predict the convergence behavior in advance in environments where multiple modulation schemes coexist.

On the other hand, learning-based \ac{MCS} selection has been extensively studied using Q-learning-based approaches~\cite{Bruno2014,Mota2019}, where a Q-table is iteratively updated to maximize the expected long-term reward for each \ac{MCS}.
However, because the Q-table size grows exponentially with the state dimension, such methods become impractical for large-scale or continuous state spaces, restricting their applicability to small state–action spaces.
To alleviate this scalability issue, \ac{DQN}-based methods employing \acp{DNN} to approximate the Q-function have been proposed~\cite{Luong2019,Zhang2019,Ye2023,Liao2023}.
While these methods can handle high-dimensional continuous state spaces without explicitly storing a Q-table, they also introduce inherent drawbacks, including increased learning complexity, sensitivity to channel variations and hyperparameters, and a lack of theoretical guarantees on convergence and stability.
Moreover, their focus on single-link adaptation and global optimization under diverse channel conditions limits their applicability to large-scale systems with stringent latency constraints.

In contrast, the approach proposed in this study is fundamentally different.
By leveraging instantaneous \ac{CSI} and average received \ac{SNR} as key features and significantly reducing the search space through offline learning, the proposed method enables fast and accurate \ac{MI} prediction and instantaneous channel-aware \ac{MCS} selection even in massive \ac{MU-MIMO-OFDM} systems.
The approach can be regarded as offloading the complex processing task to the offline domain.
Furthermore, the \ac{VDB} supports efficient updates by adding or removing data entries as long-term channel statistics change, eliminating the need for retraining.

\section{SYSTEM MODEL}
\label{sec: system model}

\subsection{Signal Model}
\label{subsec:SM}

Consider an uplink massive \ac{MU-MIMO-OFDM} system conforming to the \ac{5G NR} specification, as illustrated in Fig. \ref{fig: SystemConfig}.
Each \ac{UE} has one \ac{TX} antenna, and $M$ \acp{UE} perform spatial multiplexing to a \ac{BS} equipped with $N$ \ac{RX} antennas arranged in a \ac{URA}.
Based on the acquired \ac{CSI}, the \ac{BS} performs resource allocation and \ac{MCS} selection for uplink transmission and feeds back the results to each \ac{UE}.
Each \ac{UE} encodes and modulates its information bit stream using the selected \ac{MCS}, then transmits the signal via \acs{OFDM} over the allocated time–frequency resources.
%
%
In \ac{5G NR}~\cite{5GNR_OFDM_RB}, the smallest \ac{TB} is a \ac{RB} consisting of $14$ \acs{OFDM} symbols (time) $\times$ $12$ subcarriers (frequency).
In this work, $M$ \acp{UE} are assumed to perform spatial multiplexing transmission over time-frequency resources consisting of $R$ \acs{OFDM} symbols and $L$ subcarriers, as determined by resource allocation; thus, we focus on the transmission of a single \ac{TB} containing $R \times L$ symbols.
The cyclic prefix is assumed to be ideally inserted and removed.
At the receiver, frequency-domain \acs{MIMO} signal detection is performed via \ac{MUD}, followed by demodulation and channel decoding.

Let $x_m[r, \ell]$ denote the frequency-domain symbol transmitted by the $m$-th \ac{UE} at discrete time $r$ on the subcarrier $\ell$.
Each symbol is drawn from the \ac{QAM} constellation $\mathcal{X}$ with average energy $E_{\mathrm{s}}$ and modulation order $Q \triangleq |\mathcal{X}|$.
Denoting the \ac{TX} vector spatially multiplexed on the $[r, \ell]$-th time-frequency resource by $\bm{x}[r,\ell] \triangleq \left[x_1[r,\ell], \ldots, x_m[r,\ell], \ldots , x_M[r,\ell]\right]^{\mathsf{T}} \in \mathbb{C}^{M \times 1}$, the corresponding \ac{RX} vector can be expressed as
\begin{eqnarray}
    \bm{y}[r,\ell] 
    \!\!\!&\triangleq&\!\!\!
    [y_1[r,\ell], \ldots, y_n[r,\ell], \ldots, y_N[r,\ell]]^{\mathsf{T}} \in \mathbb{C}^{N \times 1} \nonumber \\
    \!\!\!&=&\!\!\!
    \bm{H}[\ell] \bm{x}[r,\ell] + \bm{z}[r,\ell],
    \label{eq:y}
\end{eqnarray}
where $\bm{H}[\ell] \triangleq [\bm{h}_1[\ell], \ldots, \bm{h}_m[\ell], \ldots, \bm{h}_M[\ell]] \in \mathbb{C}^{N \times M}$ denotes the frequency-domain \acs{MIMO} channel matrix of the $\ell$-th subcarrier, with $\bm{h}_m[\ell]\in\mathbb{C}^{N\times 1}$ representing its $m$-th column.
The channel is assumed to be constant over one time slot ($1 \leq r \leq R$), and $\bm{z}[r,\ell] \triangleq [z_1[r,\ell], \ldots, z_n[r,\ell], \ldots, z_N[r,\ell]]^{\mathsf{T}} \in \mathbb{C}^{N \times 1}$ is the \ac{AWGN} vector, each entry of which follows $\mathcal{CN}(0,N_0)$ with $N_0$ denoting the noise power spectral density.

Since the \ac{MCS} for each \ac{UE} must be determined prior to actual transmission, the \ac{BS} predicts the achievable demodulator output \ac{MI} under \ac{MUD} using the estimated \ac{CSI} $\bm{H}[\ell], \forall \ell$, and the average received \ac{SNR} $\rho \triangleq M E_\mathrm{s}/N_0$, without relying on the \ac{RX} signal $\bm{y}[r,\ell],\forall r,\forall \ell$.
For simplicity, both the \ac{CSI} $\bm{H}[\ell], \forall \ell$ and average received \ac{SNR} $\rho$ are assumed to be perfectly estimated at the \ac{BS}, and the assigned \ac{MCS} index is assumed to be notified to each \ac{UE} without error.

\subsection{EP-based MUD Algorithm}
\label{subsec:MUD}

\begin{algorithm}[!t]
\caption{EP-based MUD algorithm}
\label{alg: EP}
\hrulefill
\begin{algorithmic}[1]
\Require{$\bm{y}\in \mathbb{C}^{N \times 1}, \bm{H} \in \mathbb{C}^{N \times M}, T$ (Num. of iterations)}
\Ensure{$q_{m,\mathrm{A} \to \mathrm{B}}^{(T)} (\chi), \forall m, \forall \chi \in \mathcal{X}$}
\vspace{-1ex}
\Statex \hspace{-4ex}\hrulefill

\noindent {/*\raisebox{3pt}[-5pt][0pt]{ --------------- }} Initialization {\raisebox{3pt}[-5pt][0pt]{ --------------- }}*/

\State $\bm{x}_{\mathrm{B} \to \mathrm{A}} ^{(1)} = \bm{0} \in \mathbb{C}^{M \times 1}, \bm{V}_{\mathrm{B} \to \mathrm{A}} ^{(1)} = E_{\mathrm{s}}\bm{I}_M$

\noindent {/*\raisebox{3pt}[-5pt][0pt]{ ------------------ }} Iteration {\raisebox{3pt}[-5pt][0pt]{ ----------------- }}*/

\For{$t=1\ \mathrm{to}\ T$}

// Module A

\State $\bm{\varPsi}^{(t)} = \left( N_0 \bm{I}_N + \bm{H}\bm{V}_{\mathrm{B} \to \mathrm{A}}^{(t)} \bm{H}^{\mathsf{H}} \right)^{-1}$
\State $\bm{\varXi}^{(t)} = \left(\mathrm{diag} \left[ 
\bm{h}_1^{\mathsf{H}}
\bm{\varPsi}^{(t)}\bm{h}_1
,\ldots,
\bm{h}_M^{\mathsf{H}}
\bm{\varPsi}^{(t)}\bm{h}_M
\right]\right)^{-1}$
\State $\bm{x}_{\mathrm{A} \to \mathrm{B}}^{(t)} = \bm{x}_{\mathrm{B} \to \mathrm{A}}^{(t)} + \bm{\varXi}^{(t)} \bm{H}^{\mathsf{H}} \bm{\varPsi}^{(t)} \left(\bm{y} - \bm{H}\bm{x}_{\mathrm{B} \to \mathrm{A}}^{(t)}\right)$
\State $\bm{V}_{\mathrm{A} \to \mathrm{B}}^{(t)} = \bm{\varXi}^{(t)} - \bm{V}_{\mathrm{B} \to \mathrm{A}} ^{(t)}$
\State $\forall m, \forall \chi\in\mathcal{X}: q_{m,\mathrm{A} \to \mathrm{B}}^{(t)} (\chi) =  \exp{\left( -\frac{ \left| \chi - x_{m, \mathrm{A} \to \mathrm{B}}^{(t)} \right|^2 }{ v_{m, \mathrm{A} \to \mathrm{B}}^{(t)} } \right)}$

// Module B

\State $\forall m : x_{m, \mathrm{B}}^{(t)} = \frac{\sum_{\chi \in \mathcal{X}} \chi \cdot q_{m,\mathrm{A} \to \mathrm{B}} ^{(t)} (\chi)   }{\sum_{\chi' \in \mathcal{X}} q_{m,\mathrm{A} \to \mathrm{B}} ^{(t)} (\chi') }$
\State $\forall m :  v_{m, \mathrm{B}}^{(t)} = \frac{\sum_{  \chi \in \mathcal{X}} |\chi|^2\cdot q_{m,\mathrm{A} \to \mathrm{B}} ^{(t)} (\chi)   }{\sum_{\chi' \in \mathcal{X}} q_{m,\mathrm{A} \to \mathrm{B}} ^{(t)} (\chi') } - \left| x_{m, \mathrm{B}}^{(t)} \right|^2$
\State $\forall m :  \frac{1}{\overline{v}_{m, \mathrm{B} \to \mathrm{A}}^{(t)}} = \frac{1}{v_{m, \mathrm{B}}^{(t)}} - \frac{1}{v_{m, \mathrm{A} \to \mathrm{B}}^{(t)}}$
\State $\forall m :  \overline{x}_{m, \mathrm{B} \to \mathrm{A}}^{(t)} = \overline{v}_{m, \mathrm{B} \to \mathrm{A}}^{(t)} \cdot \left( \frac{x_{m, \mathrm{B}}^{^{(t)}}}{v_{m, \mathrm{B}}^{(t)}} - \frac{x_{m, \mathrm{A} \to \mathrm{B}}^{(t)}}{v_{m, \mathrm{A} \to \mathrm{B}}^{(t)}} \right)$
\State $\overline{\bm{x}}_{\mathrm{B} \to \mathrm{A}}^{(t)} = \left[\overline{x}_{1, \mathrm{B} \to \mathrm{A}}^{(t)}, \ldots, \overline{x}_{m, \mathrm{B} \to \mathrm{A}}^{(t)}, \ldots, \overline{x}_{M, \mathrm{B} \to \mathrm{A}}^{(t)}\right]^{\mathsf{T}}$
\State $\overline{\bm{V}}_{\mathrm{B} \to \mathrm{A}}^{(t)} = \mathrm{diag} \left[\overline{v}_{1, \mathrm{B} \to \mathrm{A}}^{(t)}, \ldots, \overline{v}_{m, \mathrm{B} \to \mathrm{A}}^{(t)}, \ldots, \overline{v}_{M, \mathrm{B} \to \mathrm{A}}^{(t)}\right]$
\State $\bm{x}_{\mathrm{B} \to \mathrm{A}}^{(t + 1)} = \alpha \overline{\bm{x}}_{\mathrm{B} \to \mathrm{A}}^{(t)} + (1 - \alpha)\bm{x}_{\mathrm{B} \to \mathrm{A}}^{(t)}$
\State $\bm{V}_{\mathrm{B} \to \mathrm{A}}^{(t + 1)} = \alpha \overline{\bm{V}}_{\mathrm{B} \to \mathrm{A}}^{(t)} + (1 - \alpha)\bm{V}_{\mathrm{B} \to \mathrm{A}}^{(t)}$
\EndFor
\end{algorithmic}
\end{algorithm}

The pseudocode of the \ac{EP}-based \ac{MUD} algorithm, designed following~\cite{Minka2013,Cespedes2014,Rangan2019,Takeuchi2020}, is given in Algorithm \ref{alg: EP}.
For clarity, the qualifier $[r, \ell]$ is omitted since \ac{MUD} is performed independently for each time–frequency index.
The following notation is used: $\bm{x}_{\zeta} \triangleq \left[ x_{1, \zeta}, \ldots, x_{m, \zeta}, \ldots, x_{M, \zeta} \right]^{\mathsf{T}}$, $\bm{V}_{\zeta} \triangleq \mathrm{diag} \left[ v_{1, \zeta}, \ldots, v_{m, \zeta}, \ldots, v_{M, \zeta} \right]$, where $\zeta \in \{ \mathrm{A} \to \mathrm{B}, \mathrm{B} \to \mathrm{A} \}$.
The \ac{EP} detector comprises module A, which applies soft interference cancellation and signal separation using an \ac{LMMSE} filter, and module B, which computes the conditional expectation—\textit{i.e.}, the general \ac{MMSE} solution—based on the output of module A.
Detection accuracy improves progressively through the exchange of extrinsic information between the two modules.
Further algorithmic details can be found in~\cite{Minka2013,Cespedes2014,Rangan2019,Takeuchi2020}.
The number of iterations is denoted by $T$, and $(\cdot)^{(t)}$ indicates the iteration index.
The parameter $\alpha$ in lines 14 and 15 is a damping factor.
When $T=1$, Algorithm \ref{alg: EP} reduces to the conventional \ac{LMMSE} detector.

\subsection{Computation of Bit-Wise LLRs}
\label{subsec:Cal_LLR}

Continuing from the previous subsection, the qualifier $[r, \ell]$ is omitted.
Based on the message $q_{m,\mathrm{A} \to \mathrm{B}}^{(T)} (\chi)$ from module A to B in the final iteration of Algorithm \ref{alg: EP}, the \acp{LLR} of the coded bits composing the \ac{TX} symbol $x_m$ are computed.
When $x_m$ consists of $S \triangleq \log_2Q$ coded bits $c_{m,1},\ldots,c_{m,s},\ldots,c_{m,S}$, the bit-wise \ac{LLR} corresponding to $c_{m,s}$ is given by~\cite{Kobayashi2025OJCOM}
\begin{equation}
    \lambda(c_{m,s}) \triangleq \ln\left[\frac{\sum_{\chi \in \mathcal{X}|c_{s} = 1} q_{m,\mathrm{A} \to \mathrm{B}}^{(T)} (\chi)}{\sum_{\chi' \in \mathcal{X}|c_{s} = 0} q_{m,\mathrm{A} \to \mathrm{B}}^{(T)} (\chi')}\right],
    \label{eq:LLR}
\end{equation}
where $\mathcal{X}|c_{s} = c\ (c \in \{0,1\})$ denotes the set of candidate constellation points whose $s$-th bit equals $c$.

\subsection{MI Computation per TB}
\label{subsec:Cal_MI}

Let $\lambda$ denote the bit-wise \ac{LLR} corresponding to the coded bit $c$. 
The bit-wise \ac{MI} between $c$ and $\lambda$ at the demodulator output \acp{LLR} can then be expressed as~\cite{ten2001}
\begin{eqnarray}
    &&\!\!\!\!\!\!\!\!\!\!\!\!\!\!\!\!
    I(\lambda; c) = \sum_{c \in \{0, 1\}} p_{\mathsf{c}}(c) \nonumber \\
    &&\!\!\!\!\!\!\!\!\!\!\!\!\!\!\!\!
    \int_{-\infty}^{\infty} \!\!p_{\mathrm{\lambda} \mid \mathsf{c}}(\lambda \mid c) \log_2 \left( \frac{p_{\mathrm{\lambda} \mid \mathsf{c}}(\lambda \mid c)}{\sum_{c' \in \{0, 1\}} p_{\mathsf{c}}(c') p_{\mathrm{\lambda} \mid \mathsf{c}'}(\lambda \mid c')} \right) d \lambda. 
    \label{eq:bitMI}
\end{eqnarray}
With the qualifier $[r, \ell]$, the detector output \acp{LLR} for the coded bits composing $x_m[r, \ell]$ are denoted by $\lambda(c_{m, s})[r, \ell], \forall s,$ as given in \eqref{eq:LLR}.
The total number of coded bits in one \ac{TB} is $N_\mathrm{TB} = R \times L \times S$.
Using all corresponding \acp{LLR}, the demodulator output \ac{MI} in \eqref{eq:bitMI} can be approximated as~\cite{Ibi2007}
\begin{subequations}
\begin{equation}
    I_{m,\mathrm{TB}}
    \approx
    1 - 
    \frac{1}{N_{\mathrm{TB}}}
    \sum_{r=1}^R
    \sum_{\ell=1}^L
    \sum_{s=1}^S
    \eta_{m,s}[r,\ell],
\end{equation}
\begin{equation}
    \eta_{m,s}[r,\ell]
    \triangleq
    \log_2\left(1 + \exp\left[-(2c_{m,s}[r,\ell] - 1)\lambda(c_{m,s})[r,\ell]\right]\right),
\end{equation}
\label{eq:MI_LLR}%
\end{subequations}
where $c_{m,s}[r,\ell]$ denotes the coded bit corresponding to $\lambda(c_{m,s})[r,\ell]$, which consists of the \ac{TX} symbol $x_m[r,\ell]$ in \eqref{eq:y}.

A key point to note is that \ac{MCS} selection must be completed prior to transmission; thus, the demodulator output \ac{MI} $I_{m,\mathrm{TB}}$ in \eqref{eq:MI_LLR} must be predicted solely based on the estimated \ac{CSI} and average received \ac{SNR}.
For determining the \ac{MCS}, the predicted \ac{MI} is then converted to the decoder output \ac{BLER} according to the decoding characteristics.
It is only through this two-stage process-\ac{MI} prediction followed by \ac{BLER} conversion-that the \ac{MCS} can be appropriately selected.

In the following section, we present the \ac{MI}-based \ac{MCS} selection framework, which is one of the main contributions of this work, using \ac{LMMSE}-based \ac{MUD} as a representative example.

\section{MI-based MCS Selection}
\label{sec: MI-based MCS Selection}

\begin{figure*}[t]
\centering
\includegraphics[width=2.0\columnwidth,keepaspectratio=true]{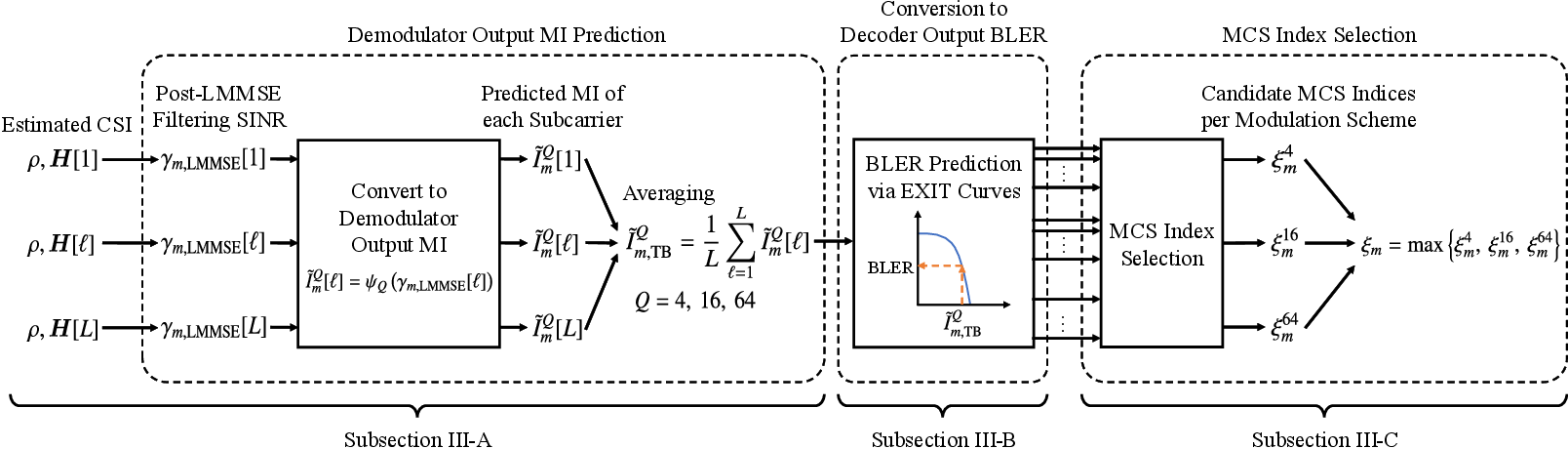}
\caption{Block diagram of the proposed \ac{MI}-based \ac{MCS} selection process incorporating the analytical \ac{MI} prediction method.}
\vspace{-3mm}
\label{fig: MI-based MCS selection analytical}
\end{figure*}

Fig. \ref{fig: MI-based MCS selection analytical} illustrates the block diagram of the \ac{MI}-based \ac{MCS} selection procedure using the analytical \ac{MI} prediction method.
In this scheme, the \ac{BS} determines the appropriate \ac{MCS} for each \ac{UE} from the estimated \ac{CSI} and the averaged received \ac{SNR}.
The process comprises three steps: (i) predicting the demodulator output (\textit{i.e.}, post-\ac{MUD}) \ac{MI} for all modulation schemes predefined in the \ac{MCS} table, (ii) converting each predicted post-\ac{MUD} \ac{MI} to the corresponding post-decoding \ac{BLER} according to the specific decoding characteristics, and (iii) selecting the \ac{MCS} index that maximizes throughput while satisfying the reference \ac{BLER}.
Note that since \ac{MI} depends on the modulation scheme, the step (i) must be performed individually for each scheme.
The detailed procedure is presented below, where the \ac{LMMSE} filter is employed as \ac{MUD}.

\subsection{Analytical \ac{MI} Prediction for \ac{LMMSE} Filter Output}
\label{subsec:LMMSE_MI}

\begin{figure}[t]
\vspace{-2mm}
\centering
\includegraphics[width=0.9\columnwidth,keepaspectratio=true]{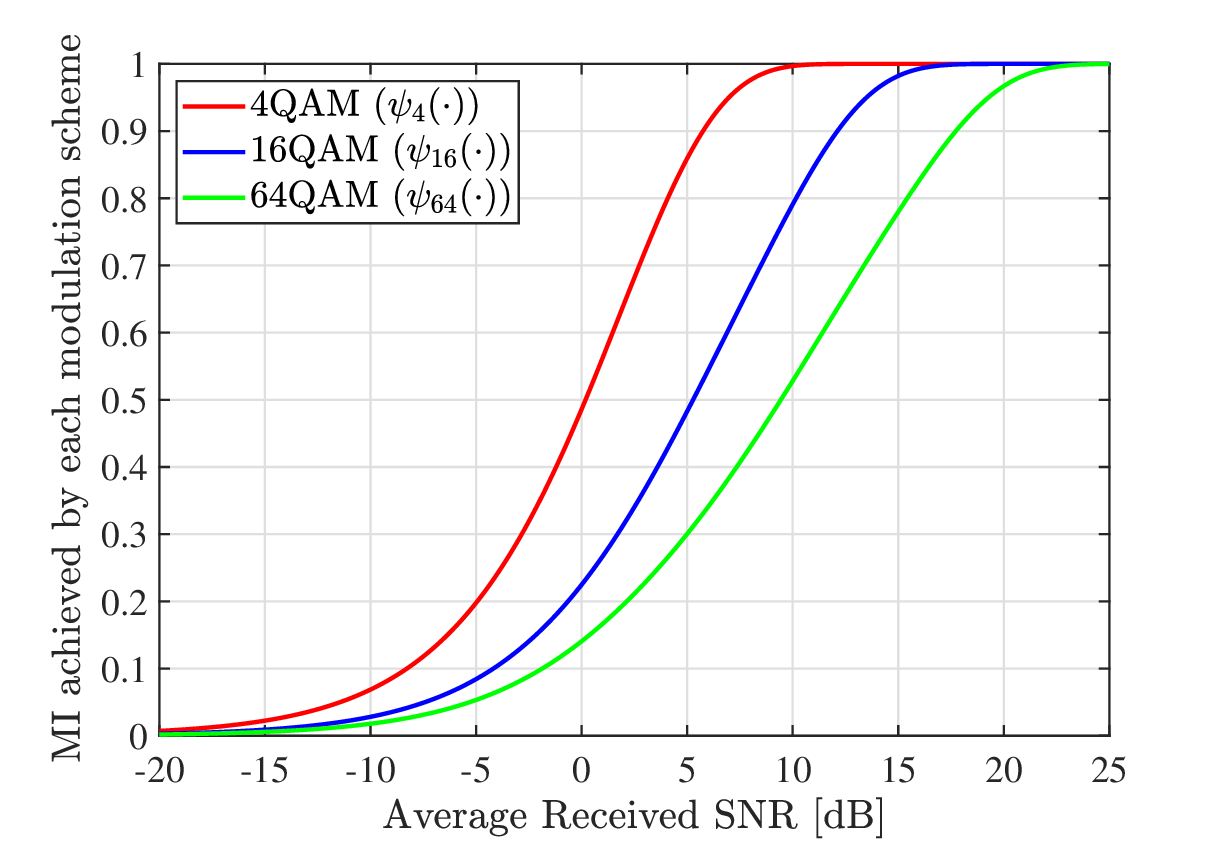}
\vspace{-1mm}
\caption{Average received SNR versus MI in an \ac{AWGN} channel.}
\label{fig: AWGN SNR vs inMI}
\vspace{-3mm}
\end{figure}

When employing the \ac{LMMSE} filter as \ac{MUD}, the post-\ac{MUD} \ac{SINR} can be derived analytically.
For detecting $x_m[r,\ell],\forall r,$ from the \ac{RX} signal $\bm{y}[r,\ell],\forall r,$ in \eqref{eq:y}, the \ac{SINR} at the \ac{LMMSE} filter output can be expressed in closed-form as~\cite{Paulraj2008}
\begin{equation}
    \gamma_{m, \mathrm{LMMSE}}[\ell] = \frac{1}{\left[\left(\rho \bm{H}^{\mathsf{H}}[\ell]\bm{H}[\ell] + \bm{I}_{M} \right)^{-1} \right]_{m,m}} - 1.
    \label{eq:SINR}
\end{equation}
Next, by modeling the \ac{LMMSE} filter output as the output of an \ac{AWGN} channel with the same \ac{SNR} as in \eqref{eq:SINR}, the post-\ac{MUD} \ac{SINR} for each subcarrier can be converted to the predicted demodulator output \ac{MI} as $\tilde{I}_{m}^Q[\ell] = \psi_{Q}\left(\gamma_{m, \mathrm{LMMSE}}[\ell]\right)$, where $\psi_{Q}(\cdot)$, with $Q\in\left\{4,16,64\right\}$, is the conversion function shown in Fig. \ref{fig: AWGN SNR vs inMI}.
By averaging over all $L$ subcarriers, the predicted demodulator output \ac{MI} $I_{m,\mathrm{TB}}$ in \eqref{eq:MI_LLR} can be expressed as
\begin{equation}
    \tilde{I}_{m,\mathrm{TB}}^Q = \frac{1}{L} \sum_{\ell=1}^L \tilde{I}_{m}^Q[\ell],\quad \forall Q.
    \label{eq:PreMI_LMMSE}
\end{equation}
For \ac{MCS} selection, the post-\ac{MUD} \acp{MI} are first computed using \eqref{eq:PreMI_LMMSE} for all modulation orders.
Each predicted \ac{MI} is then converted to the corresponding decoder output \ac{MI} according to the decoding characteristics, and the highest \ac{MCS} index meeting the reference \ac{BLER} is selected.

\begin{figure}[t]
\centering
\includegraphics[width=0.96\columnwidth,keepaspectratio=true]{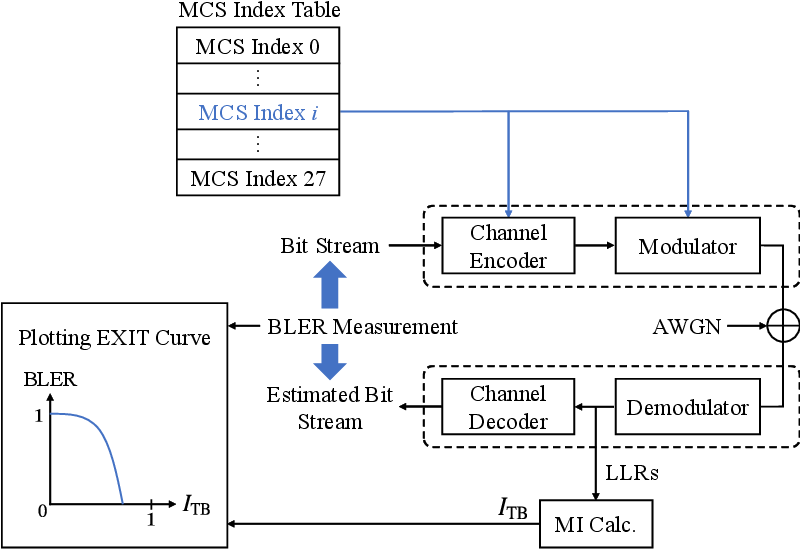}
\caption{Block diagram for measuring the relationship between the demodulator output \ac{MI} and the decoder output \ac{BLER} in an \ac{AWGN} channel.}
\label{fig: AWGN inMI vs BLER}
\vspace{-3mm}
\end{figure}

\begin{figure*}[!t]
\begin{center}
    \subfloat[$4$QAM.]{
    \includegraphics[width=0.66\columnwidth,keepaspectratio=true]{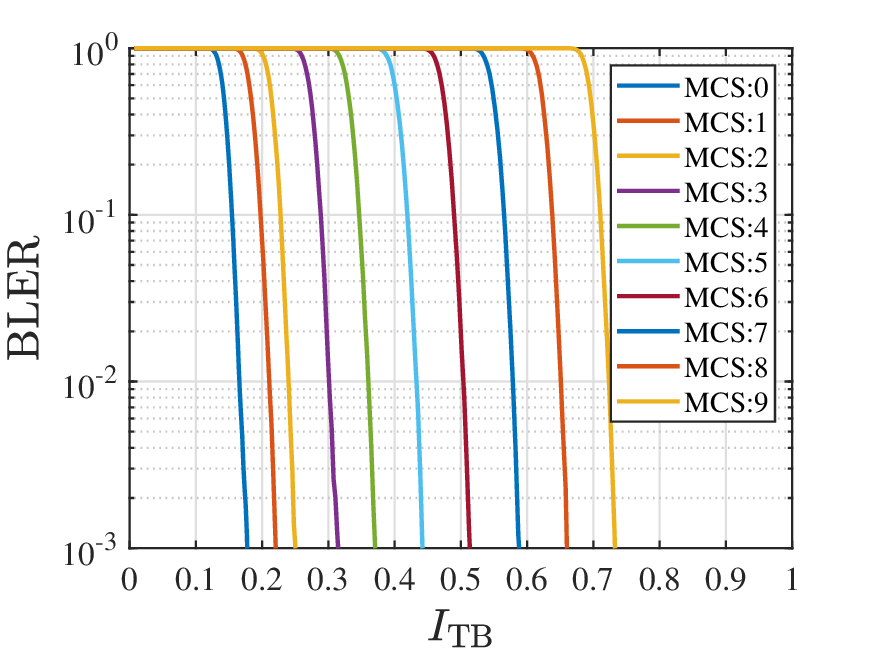}
    }
    \subfloat[$16$QAM.]{
    \includegraphics[width=0.66\columnwidth,keepaspectratio=true]{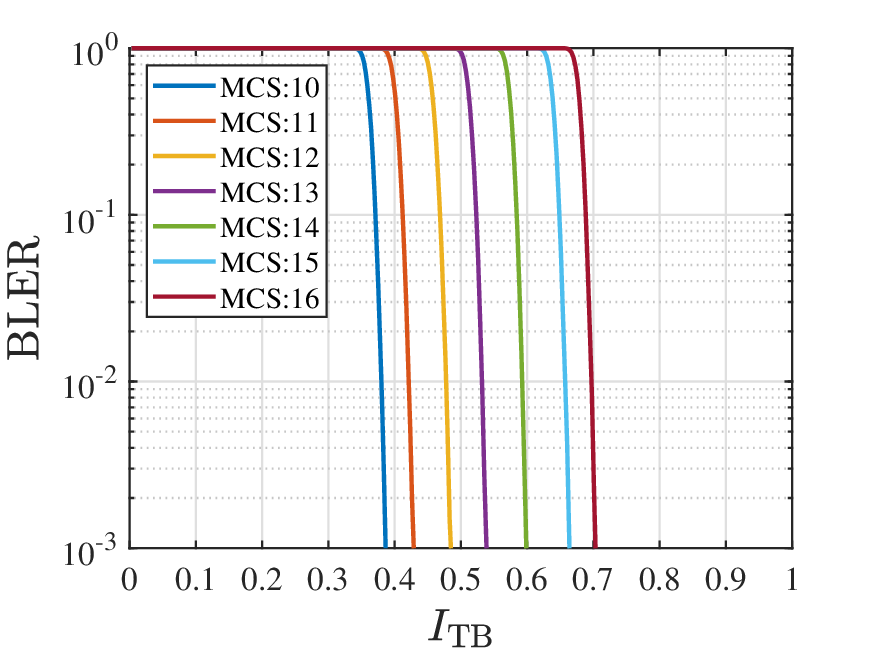}
    }
    \subfloat[$64$QAM.]{
    \includegraphics[width=0.66\columnwidth,keepaspectratio=true]{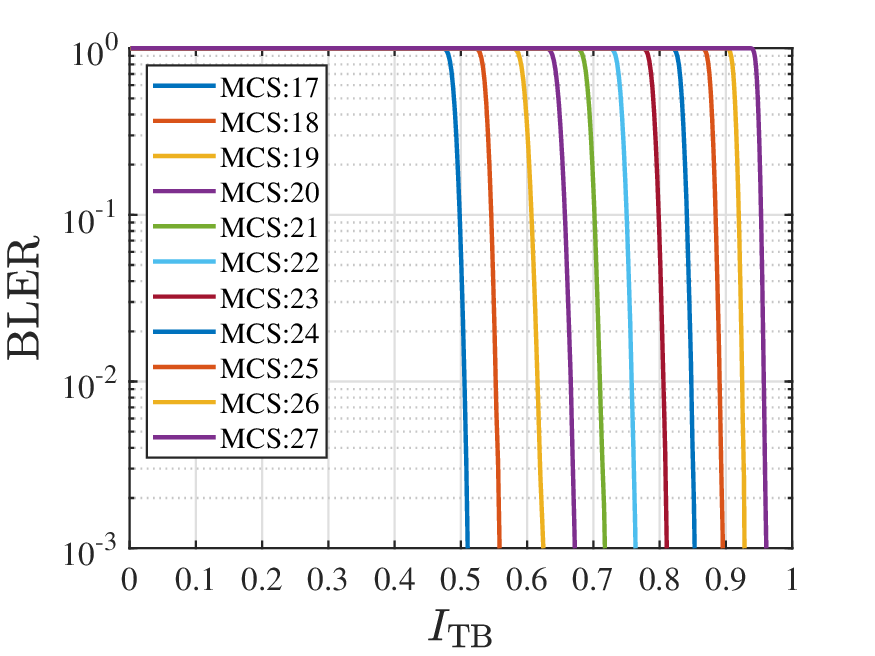}
    }
    \caption{EXIT curves showing decoder input \ac{MI} $I_{m,\mathrm{TB}}$ versus \ac{BLER} for each \ac{MCS}.}
    \label{fig: EXIT curve}
\end{center}
\vspace{-10mm}
\end{figure*}

\begin{figure*}[!t]
\begin{center}
    \subfloat[LMMSE detection, $4$QAM.]{
    \includegraphics[width=0.66\columnwidth,keepaspectratio=true]{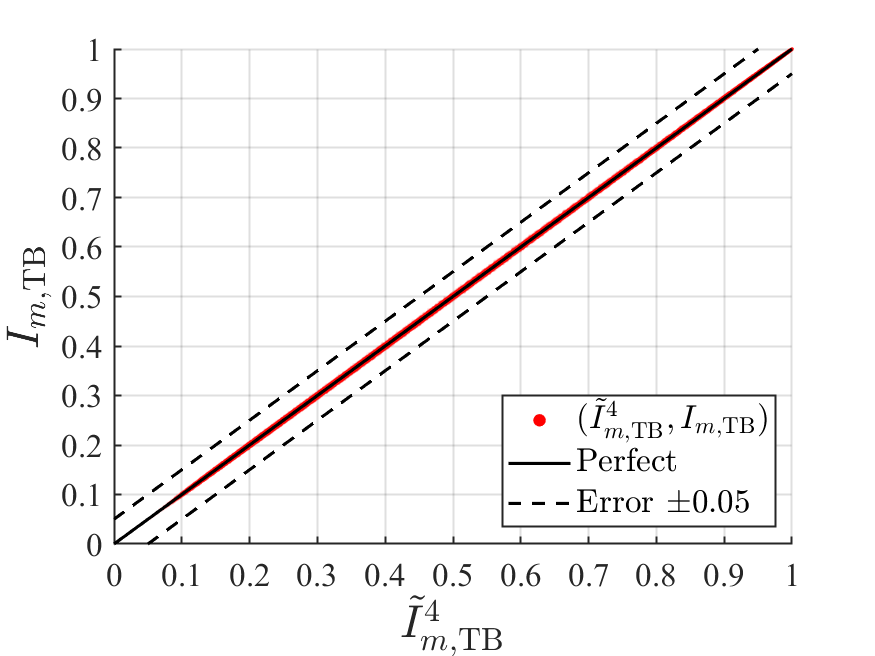}
    }
    \subfloat[LMMSE detection, $16$QAM.]{
    \includegraphics[width=0.66\columnwidth,keepaspectratio=true]{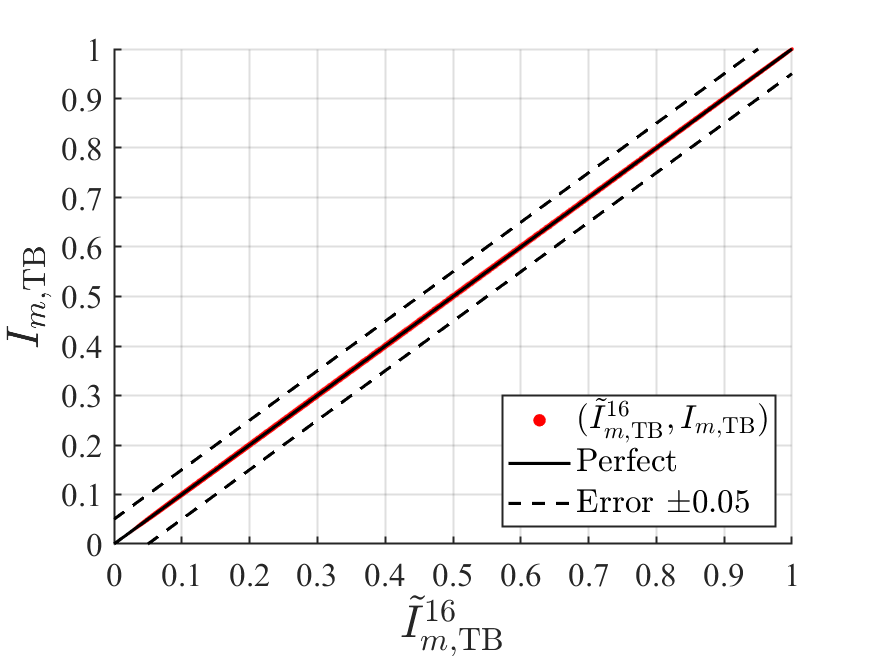}
    }
    \subfloat[LMMSE detection, $64$QAM.]{
    \includegraphics[width=0.66\columnwidth,keepaspectratio=true]{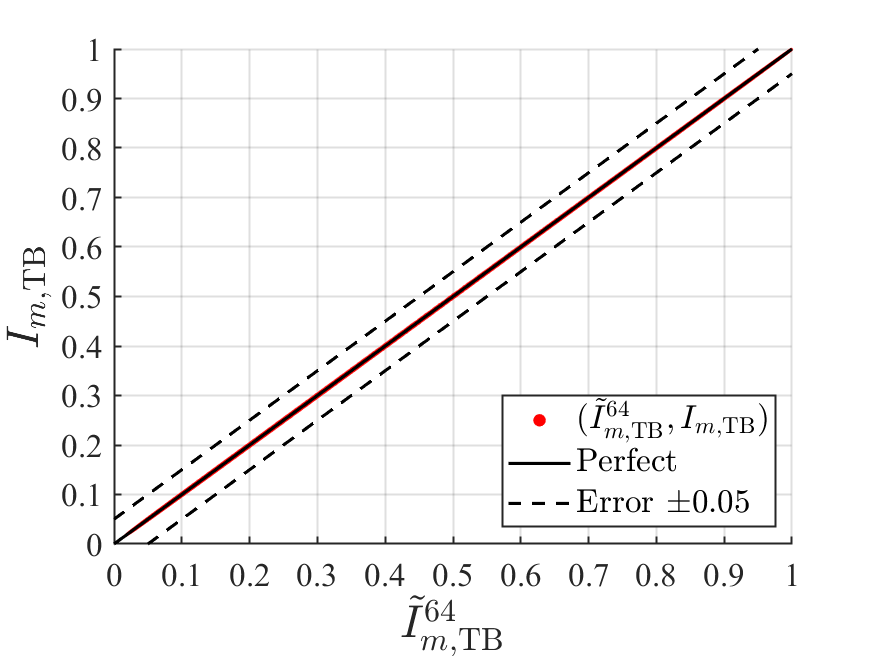}
    }
    \\
    \vspace{-4mm}
    \subfloat[EP detection, $4$QAM.]{
    \includegraphics[width=0.66\columnwidth,keepaspectratio=true]{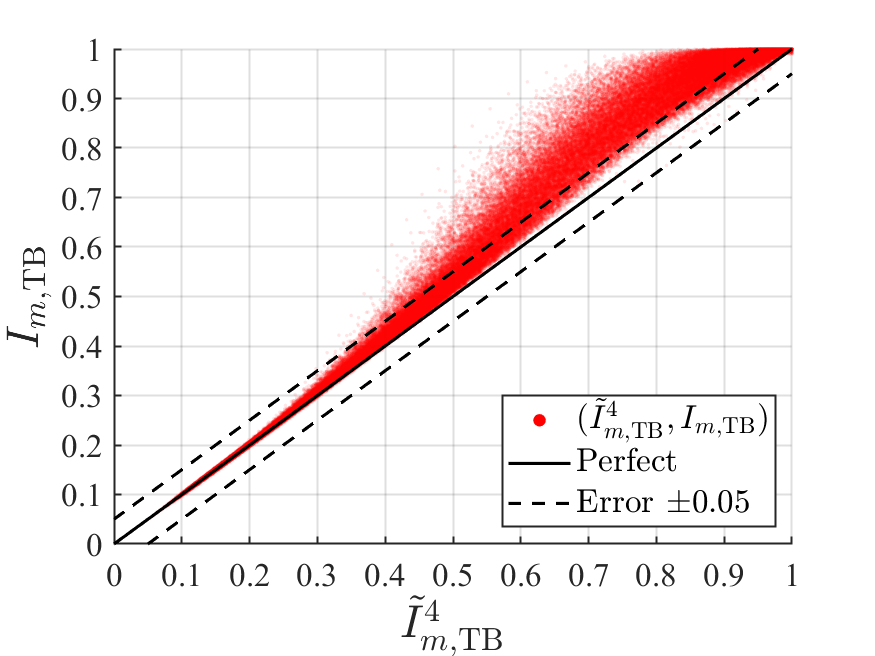}
    }
    \subfloat[EP detection, $16$QAM.]{
    \includegraphics[width=0.66\columnwidth,keepaspectratio=true]{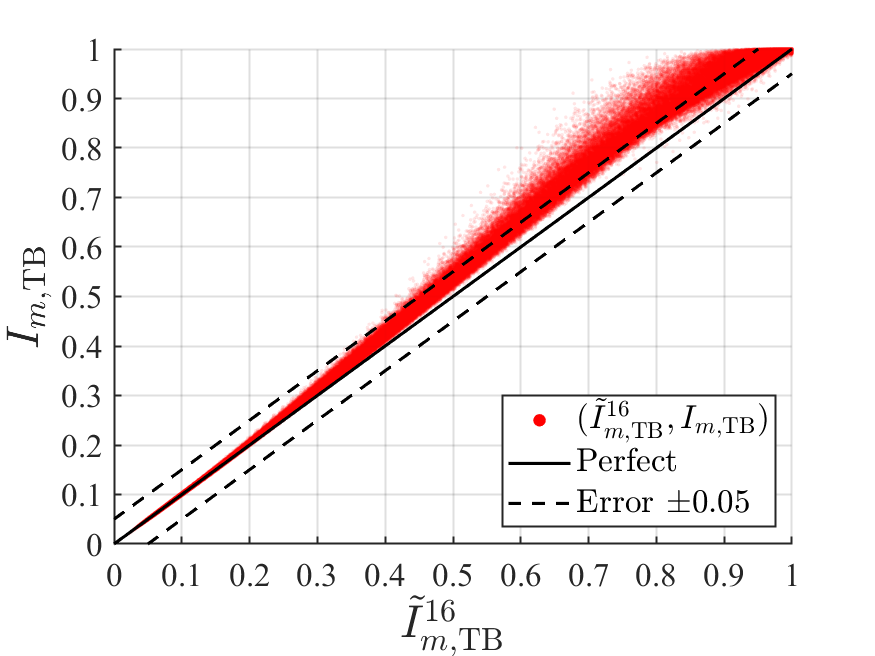}
    }
    \subfloat[EP detection, $64$QAM.]{
    \includegraphics[width=0.66\columnwidth,keepaspectratio=true]{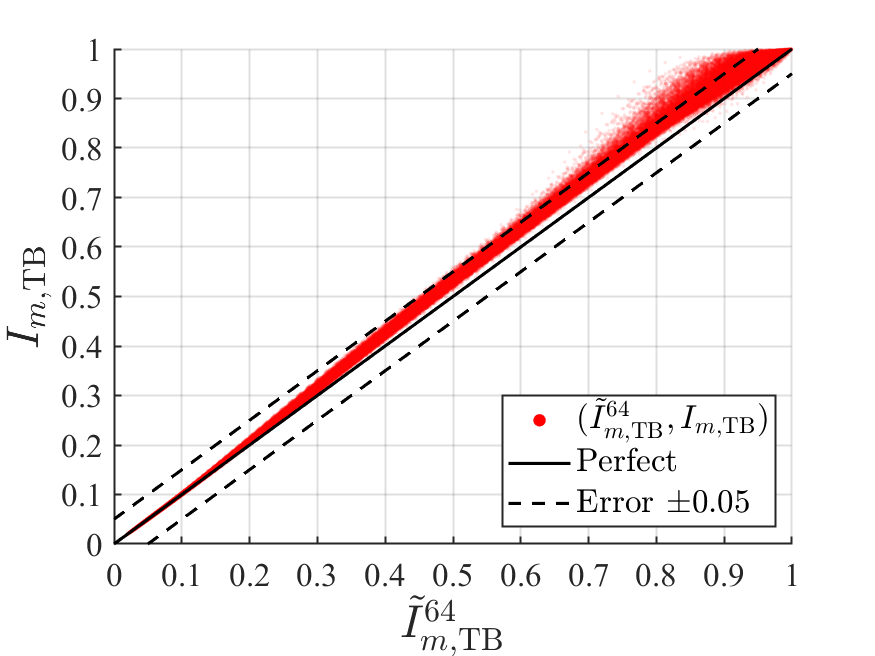}
    }
    \caption{Scatter plots of the measured \ac{MI} $I_{m,\mathrm{TB}}$ versus the predicted \ac{MI} $\tilde{I}_{m,\mathrm{TB}}^Q$ based on the \ac{LMMSE} detector.}
    \label{fig:res_Mean}
\end{center}
\vspace{-7mm}
\end{figure*}

\subsection{Conversion to Decoder Output \ac{BLER}}
\label{subsec:ConversionMI}

The decoding characteristics are modeled using \ac{EXIT} curves~\cite{Ashikhmin2004,Sharon2006,Ibi2007}.
Traditionally, these curves are obtained by generating ideal \ac{LLR} sequences that satisfy the consistency condition\footnote{The distribution of ideal \acp{LLR} satisfying the consistency condition is the Gaussian distribution with the mean $\mu$ and the variance $\sigma^2 = 2\mu$.}~\cite{Hagenauer2004} and measuring the decoder input–output \ac{MI}.
However, for higher-order modulation schemes (\textit{e.g.}, $16$\ac{QAM}, $64$\ac{QAM}), the decoder input \acp{LLR} deviate from the ideal statistics due to the non-orthogonal mapping between bits and constellation points~\cite{Kobayashi2025OJCOM}. 
This mismatch causes the \ac{MI} observed in practice to differ from that predicted by the ideal \ac{EXIT} curve.
In addition, for \ac{MCS} selection, the relevant metric is the decoder output \ac{BLER}, not \ac{MI}.
Consequently, unlike conventional \ac{EXIT} curves, which focus solely on \ac{MI}, the derived curves must capture \ac{BLER} characteristics to enable accurate \ac{MCS} selection.

\begin{table}[t]
  \caption{Simulation Parameters}
  \vspace{-2mm}
  \label{table: prm MIMO-OFDM}
  \centering
  \scalebox{0.8}{
  \begin{tabular}{c | c} 
    \hline
    Item & Value \\
    \hline
    MIMO configuration & $(N,M)=(64,16)$\\
    Center frequency & $4.7$ GHz \\
    Bandwidth & $100$ MHz \\
    Subcarrier spacing & $30$ kHz \\
    Number of subcarriers & $192$ \\
    Number of FFT point & $256$ \\
    Delay spread & $100$ ns \\
    Delay profile & CDL-B \cite{5GNR_Channel} \\
    Channel estimation & Perfect \\
    Modulation scheme & Gray-coded $4$QAM, $16$QAM, $64$QAM \\
    Channel coding scheme & LDPC \cite{5GNR_Coding} \\
    Number of EP iterations & $T = 8$ \\
    Damping Parameter & $\alpha = 0.5$ \\
    \hline
  \end{tabular}
  }
  \vspace{-4mm}
\end{table}

To ensure accurate conversion from predicted post-\ac{MUD} \ac{MI} to decoder output \ac{BLER} (post-decoding \ac{BLER}), we derive \ac{EXIT} curves using the procedure illustrated in Fig. \ref{fig: AWGN inMI vs BLER}, which reflects the actual \ac{LLR} statistics for each \ac{MCS} index:
\begin{enumerate}
    \item Simulation Setup: Generate coded symbols according to the target \ac{MCS} index.
    \item Channel Modeling: Pass the symbols through an \ac{AWGN} channel adjusted to achieve the desired input \ac{MI}.
    \item \ac{LLR} Computation: Obtain demodulator output \acp{LLR} directly from the channel output, preserving their actual statistical distribution. 
    \item Decoder Evaluation: Measure the decoder input \ac{MI} from these \acp{LLR} and determine the output \ac{BLER} by running the channel decoder.
    \item Curve Construction: Repeat over a range of input \acp{MI} to obtain the complete \ac{EXIT} curve.
\end{enumerate}

By precomputing \ac{EXIT} curves for all \ac{MCS} indices in the \ac{5G NR}–compliant \ac{MCS} table, the post-decoding \ac{BLER} for each \ac{MCS} can be predicted directly from the demodulator output \ac{MI}—eliminating the need for the actual decoding process.
These curves are generated to reflect the statistical properties of the \acp{LLR} in practical higher-order modulations.

\begin{figure*}[t]
\begin{center}
    \subfloat[LMMSE detection, $4$QAM.]{
    \includegraphics[width=0.66\columnwidth,keepaspectratio=true]{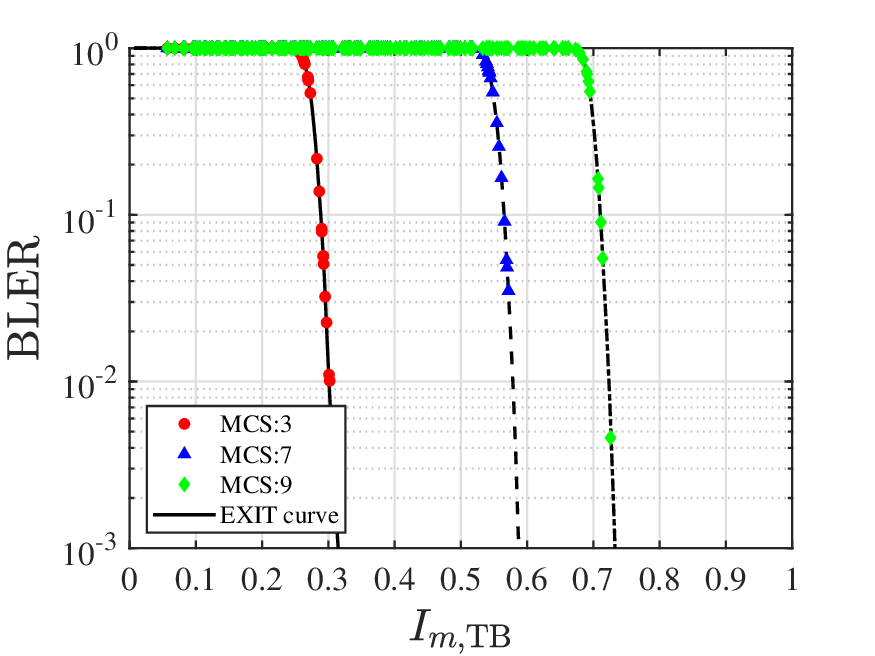}
    }
    \subfloat[LMMSE detection, $16$QAM.]{
    \includegraphics[width=0.66\columnwidth,keepaspectratio=true]{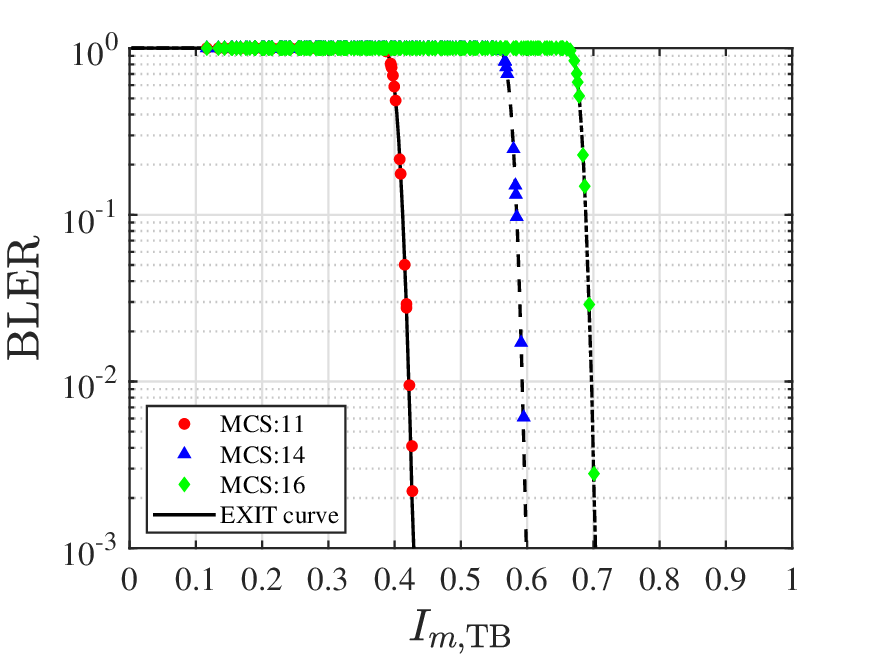}
    }
    \subfloat[LMMSE detection, $64$QAM.]{
    \includegraphics[width=0.66\columnwidth,keepaspectratio=true]{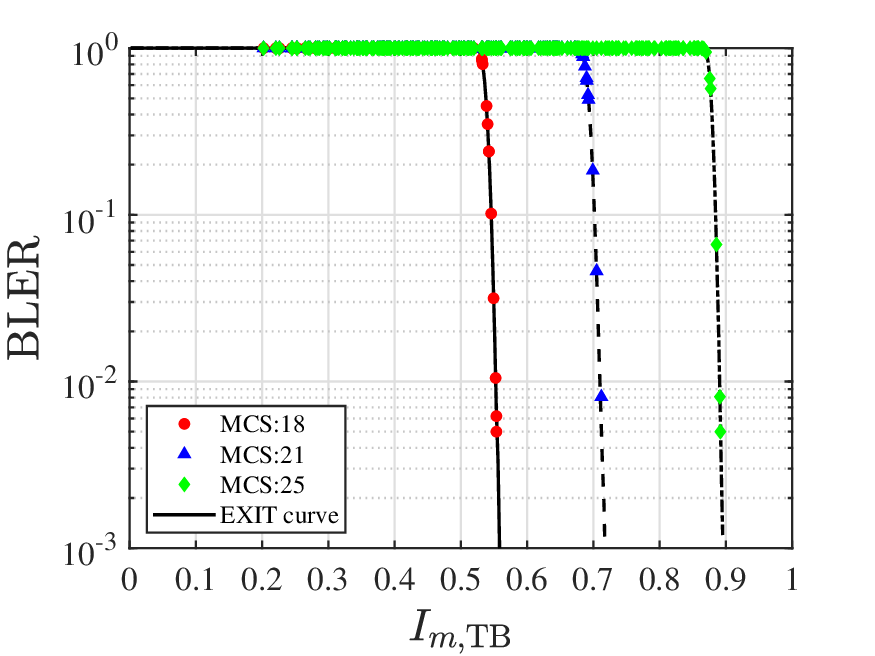}
    }
    \\
    \vspace{-3mm}
    \subfloat[EP detection, $4$QAM.]{
    \includegraphics[width=0.66\columnwidth,keepaspectratio=true]{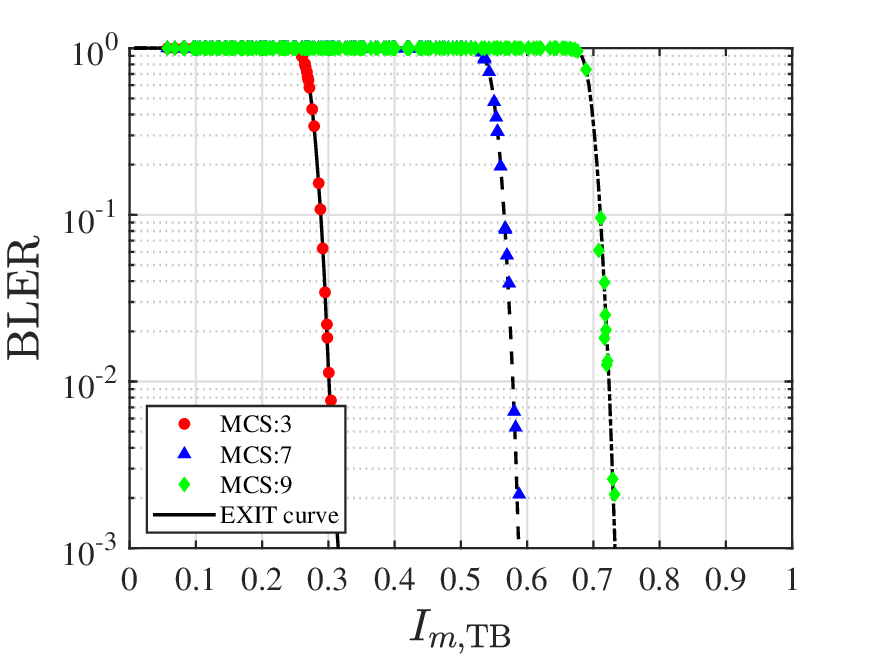}
    }
    \subfloat[EP detection, $16$QAM.]{
    \includegraphics[width=0.66\columnwidth,keepaspectratio=true]{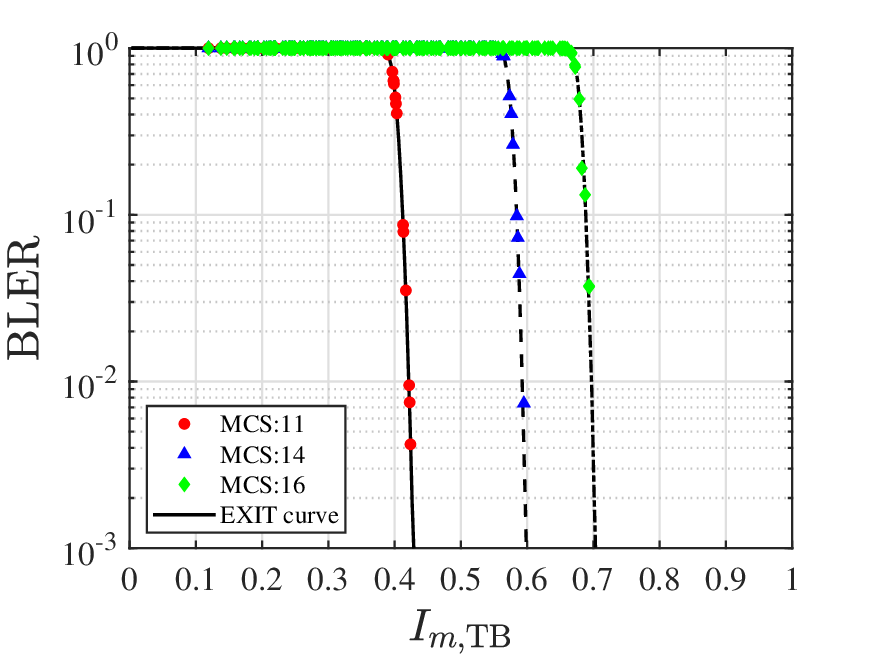}
    }
    \subfloat[EP detection, $64$QAM.]{
    \includegraphics[width=0.66\columnwidth,keepaspectratio=true]{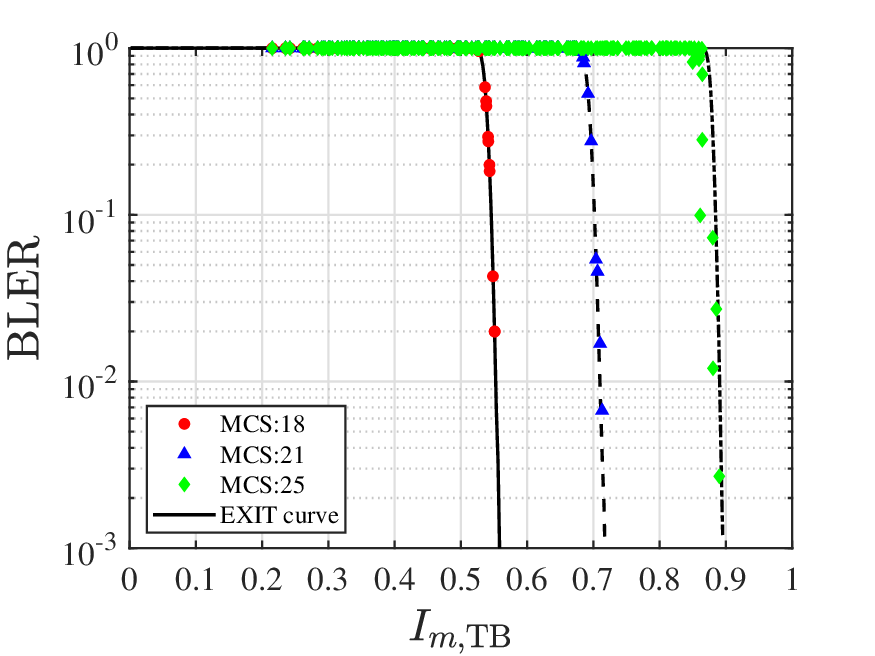}
    }
    \caption{Scatter plots of decoder input \ac{MI} $I_{m,\mathrm{TB}}$ versus \ac{BLER}.}
    \label{fig: MI-BLER}
\end{center}
\vspace{-8mm}
\end{figure*}

Fig. \ref{fig: EXIT curve} shows the resulting \ac{EXIT} curves, illustrating the relationship between the decoder input \ac{MI} $I_{\mathrm{TB}}$ and the post-decoding \ac{BLER} for each \ac{MCS} index.
The \ac{MCS} table follows the \ac{5G NR} specification~\cite{5GNR_MCS} and defines $28$ \ac{MCS} indices, grouped by modulation scheme: $4$\ac{QAM} (indices $0$–$9$), $16$\ac{QAM} (indices $10$–$16$), and $64$\ac{QAM} (indices $17$ to $27$).
\Ac{LDPC} codes~\cite{5GNR_Coding} are used for channel coding.
As the \ac{MCS} index increases, the corresponding code rate—and thus the data rate—also increases.
Figs. \ref{fig: EXIT curve} (a)–(c) correspond to $4$\ac{QAM}, $16$\ac{QAM}, and $64$\ac{QAM}, respectively.

\subsection{\ac{MCS} Index Selection}
\label{subsec:MCSIndexSelection}

Given the predicted post-\ac{MUD} \ac{MI} $\tilde{I}_{m,\mathrm{TB}}^Q$ in \eqref{eq:PreMI_LMMSE} for $Q\in\left\{4, 16, 64\right\}$, the highest \ac{MCS} index $\left\{ \xi_m^4, \xi_m^{16}, \xi_m^{64} \right\}$ that satisfies the reference \ac{BLER} is selected for each \ac{UE} using the \ac{EXIT} curves in Fig. \ref{fig: EXIT curve}. 
The final \ac{MCS} index $\xi_m$ assigned to each \ac{UE} is then determined as the maximum among these candidates:
\begin{equation}
    \xi_m = \max \left( \xi_m^4, \xi_m^{16}, \xi_m^{64} \right).
\end{equation}
This procedure underscores that in \ac{MI}-based \ac{MCS} selection, accurate prediction of the post-\ac{MUD} \ac{MI} is essential for achieving optimal performance.

\subsection{Evaluation of Post-MUD MI Prediction}
\label{subsec:EvaluationMIPrediction}

Computer simulations were performed to evaluate the accuracy of the post-\ac{MUD} \ac{MI} predicted by \eqref{eq:PreMI_LMMSE} for the \ac{LMMSE} filtering in a massive \ac{MU-MIMO-OFDM} system.
The simulation parameters are summarized in Tab. \ref{table: prm MIMO-OFDM}.
Sixteen synchronized \acp{UE} ($M=16$), each with a single \ac{TX} antenna, simultaneously transmit to a \ac{BS} equipped with $N=64$ \ac{RX} antennas.
The \ac{3GPP} \ac{CDL} channel model~\cite{5GNR_Channel} is employed.
The carrier frequency is $4.7$ GHz, the system bandwidth is $100$ MHz, and the \acs{OFDM} subcarrier spacing is $30$ kHz.
We consider one time-slot \acs{OFDM} transmission over $L=192$ subcarriers ($16$ \acp{RB}) with an \ac{FFT} size of $256$.
The number of coded bits per \ac{TB} is $N_{\mathrm{TB}} = R \times L \times S = 2688 S$.
The demodulator output \ac{MI} is computed from \eqref{eq:MI_LLR} for each $N_\mathrm{TB}$-bit codeword.
To isolate prediction accuracy, all \acp{UE} are assumed to use the same modulation scheme ($Q\in\left\{4,16,64\right\}$), and the average received \ac{SNR} $\rho$ is varied from $-10$ dB to $24$ dB in $0.5$ dB increments.

Fig.~\ref{fig:res_Mean} shows scatter plots of the measured \ac{MI} $I_{m,\mathrm{TB}}$ from \eqref{eq:MI_LLR} (vertical axis) versus the predicted \ac{MI} $\tilde{I}_{m,\mathrm{TB}}^Q$ from \eqref{eq:PreMI_LMMSE} (horizontal axis).
The solid black line represents the ideal case ($I_{m,\mathrm{TB}} = \tilde{I}_{m,\mathrm{TB}}^Q$), indicating perfect prediction, while the dashed lines ($I_{m,\mathrm{TB}} = \tilde{I}_{m,\mathrm{TB}}^Q\pm 0.05$), denote a reference \ac{PE} of $0.05$.
Figs.~\ref{fig:res_Mean}(a)-(c) present results for \ac{LMMSE}-based \ac{MUD} across all modulation schemes.
As expected, the measured samples (red dots) lie closely along the ideal line, confirming the high accuracy of the analytical \ac{MI} prediction.
This validates the accuracy of the prediction procedure described in Subsection III-A.

Next, we investigate the deviation observed when the actual \ac{MUD} employs \ac{EP}-based iterative detection, compared with predictions assuming \ac{LMMSE}-based \ac{MUD}.
As shown in Figs.~\ref{fig:res_Mean}(d)-(f) for \ac{EP}-based \ac{MUD} ($T=8$ in Algorithm \ref{alg: EP}), most samples lie above the ideal line due to the iterative gain in \ac{MI} over the \ac{LMMSE}-based prediction.

These results indicate that using \eqref{eq:PreMI_LMMSE} for \ac{MCS} selection together with \ac{EP}-based \ac{MUD} can provide highly reliable communication while substantially reducing retransmissions, and further suggest that if \ac{MI} prediction tailored for \ac{EP}-based \ac{MUD} could achieve accuracy comparable to that for \ac{LMMSE}-based \ac{MUD}, additional throughput gains could be realized through better-aligned \ac{MCS} selection.
%


\subsection{Evaluation of Post-Decoding BLER Prediction}
\label{subsec:EvaluationBLERPrediction}

Finally, using the same simulation settings as in the previous subsection, we evaluate the accuracy of converting post-\ac{MUD} \ac{MI} to post-decoding \ac{BLER} via the \ac{EXIT} curves in Fig.~\ref{fig: EXIT curve}.
Fig.~\ref{fig: MI-BLER} shows \ac{BLER} versus the measured \ac{MI} $I_{m,\mathrm{TB}}$ from \eqref{eq:MI_LLR}, with the aim of isolating the \ac{EXIT}-curve-based conversion accuracy; thus, predicted \ac{MI} is not used in this evaluation.
The black curves are the \ac{EXIT} curves for the selected \ac{MCS} indices, directly taken from Fig.~\ref{fig: EXIT curve}, while the scatter plots represent (post-\ac{MUD} \ac{MI}, post-decoding \ac{BLER}) pairs measured in actual \ac{MU-MIMO-OFDM} transmission.
For each modulation scheme, three \ac{MCS} indices from the \ac{5G NR} \ac{MCS} table~\cite{5GNR_MCS} are tested. 
Figs.~\ref{fig: MI-BLER} (a)-(c) present results with \ac{LMMSE}-based \ac{MUD} ($T = 1$ in Algorithm \ref{alg: EP}), and Figs.~\ref{fig: MI-BLER} (d)-(f) with \ac{EP}-based \ac{MUD} ($T = 8$).

In both cases, the measured samples align closely with the \ac{EXIT} curves, confirming that even with nonlinear iterative detection, each stream after \ac{MUD} behaves equivalently to an \ac{AWGN} channel and that the designed \ac{EXIT} curves accurately characterize the relationship between decoder input (\textit{i.e.}, post-\ac{MUD}) \ac{MI} and post-decoding \ac{BLER}.
This validates the accurate modeling of the decoding characteristics described in Subsection III-B.
Therefore, the \ac{EXIT} curves in Fig.~\ref{fig: EXIT curve} can serve as conversion functions from predicted post-\ac{MUD} \ac{MI} $\tilde{I}_{m,\mathrm{TB}}$ to predicted \ac{BLER}.
By preparing \ac{EXIT} curves for each code length in advance, the predicted post-\ac{MUD} \ac{MI} can be directly mapped to the expected post-decoding \ac{BLER}
for any \ac{MCS} index.

The above findings indicate that realizing the system throughput gains offered by advanced iterative \ac{MUD} requires accurate prediction of the enhanced post-\ac{MUD} \ac{MI} observed in Figs.~\ref{fig:res_Mean} (d)–(f).
In the next section, we present a prediction method specifically designed to capture this improvement.

\section{Proposed VSS-based MI Prediction}
\label{sec: Proposed Method}

\begin{figure}[!t]
\vspace{4mm}
\begin{center}
    \subfloat[Offline VDB construction phase.]{
    \includegraphics[width=1.0\columnwidth]{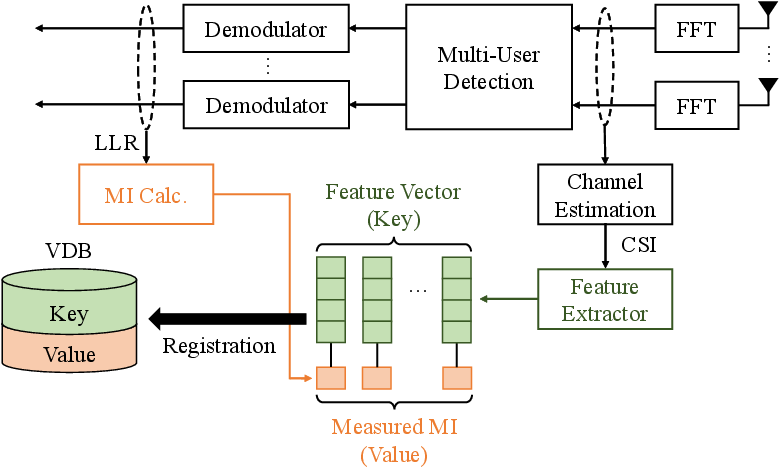}
    }
    \\
    \subfloat[Online MI prediction phase.]{
    \includegraphics[width=1.0\columnwidth]{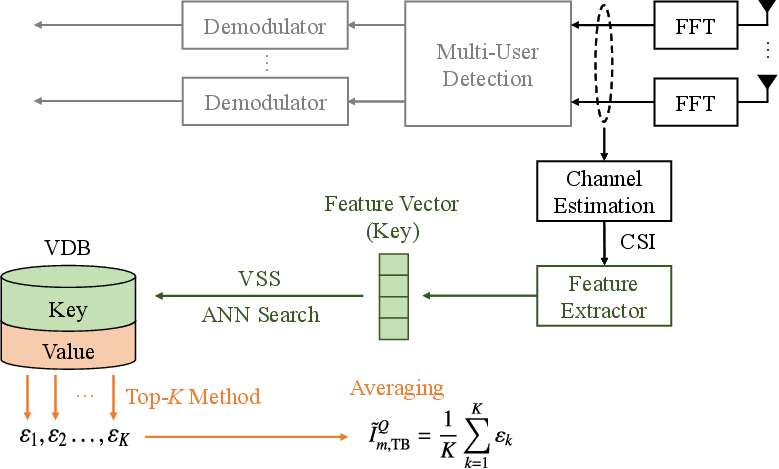}
    }
    \caption{Block diagram of \ac{MI} prediction using the \ac{VDB} and \ac{VSS}.}
    \label{fig:MIpre_VDB}
\end{center}
\vspace{-4mm}
\end{figure}

Fig. \ref{fig:MIpre_VDB} illustrates the proposed \ac{VSS}-based \ac{MI} prediction framework employing \ac{ANN} search over a \ac{VDB}, which comprises two phases: (a) an offline \ac{VDB} construction phase and (b) an online \ac{MI} prediction phase.
In the offline phase (a), a feature vector (key) is generated from the \ac{CSI} $\bm{H}$ and the average received \ac{SNR} $\rho$ and is paired with the measured \ac{MI} (value) obtained from \ac{MU-MIMO-OFDM} transmission over the corresponding wireless channel when detection is performed using the \ac{EP}-based \ac{MUD}.
These key–value pairs are stored in the \ac{VDB} in vector format.
This process is typically performed offline—either through simulations or measurements in real environments—but can also be executed online, with the \ac{VDB} being incrementally updated as new measurement data becomes available.

In the online phase (b), a key is generated in the same manner from the estimated \ac{CSI} and average received \ac{SNR}.
An \ac{ANN} search is then conducted on the \ac{VDB} to retrieve a specified number of nearest-neighbor keys, and the associated values are returned.
These values correspond to \acp{MI} achieved under similar channel conditions; thus, with an appropriately designed feature vector, highly accurate \ac{MI} predictions can be achieved.
In this study, the Faiss library~\cite{Douze2025} is adopted for \ac{ANN} search due to its efficient \ac{GPU}-based implementation.
The detailed procedures for both phases are described in the following subsections.

\subsection{Offline VDB Construction Phase}
\label{subsec:OfflineVDBconstruction}

The \ac {VDB} contains $D_{\mathrm{VDB}}$ entries, each comprising a feature vector (key) $\bm{\upsilon} \in \mathbb{R}^{D_{\mathrm{key}} \times 1}$ and its associated acquired knowledge (value) $\varepsilon\in \mathbb{R}$, stored as a pair $(\bm{\upsilon},\varepsilon)$.
%
%
In this study, keys are extracted from the \ac{CSI} $\bm{H}\in \mathbb{C}^{N \times M \times L}$ and the average received \ac{SNR} $\rho\in\mathbb{R}$ via the feature extraction function:
\begin{equation}
    \bm{\upsilon} = f\left(\bm{H},\rho\right),
    \label{eq:keyfunc}
\end{equation}
whose design critically influences prediction accuracy.

Since \ac{MCS} selection requires \ac{MI} prediction on a per-\ac{UE} basis, feature vectors are computed individually for each \ac{UE}, with $\bm{\upsilon}_m \in \mathbb{R}^{D_{\mathrm{key}} \times 1}$ denoting the vector for the $m$-th \ac{UE}, as illustrated in Fig.~\ref{fig:MIpre_VDB} (a).
The values $\varepsilon$ correspond to the measured demodulator output (post-\ac{MUD}) \ac{MI} $I_{m,\mathrm{TB}}$ in \eqref{eq:MI_LLR}, obtained from the \acp{LLR} computed after actual \acs{OFDM} transmission over wireless channel $\bm{H}$ and subsequent \ac{MUD}. 
A large set of channel realizations is generated according to a stochastic channel model, and all resulting $(\bm{\upsilon}_m, I_{m,\mathrm{TB}})$ pairs are stored in the \ac{VDB} for later \ac{ANN}-based retrieval.

\subsection{Online MI Prediction Phase}
\label{subsec:OnlineMIPrediction}

The first task is to compute the feature vector corresponding to the current channel state, following the same procedure as in the offline phase.
Let the estimated \ac{CSI} and average received \ac{SNR} be denoted by $\acute{\bm{H}} \in \mathbb{C}^{N \times M \times L}$ and $\acute{\rho} \in \mathbb{R}$, respectively.
The key for the $m$-th \ac{UE} is then generated using \eqref{eq:keyfunc} as
\begin{equation}
    \acute{\bm{\upsilon}}_m = f\left(\acute{\bm{H}},\acute{\rho}\right),
    \label{eq:find_nu}
\end{equation}
and used as a query vector for \ac{ANN} search on the \ac{VDB}.

In the \ac{GPU}-based implementation, the Faiss library~\cite{Douze2025} employs two key techniques to enable large-scale, low-latency \ac{VSS}.
The first is \ac{IVF} indexing~\cite{Sivic2003}, which partitions the database into $K_{\mathrm{IVF}}$ clusters using k-means clustering and stores only the representative vector (\textit{centroid}) of each cluster, with each data point associated with its corresponding centroid.
The second is \ac{PQ}~\cite{Jégou2011}, which decomposes each $D_\mathrm{key}$-dimensional key into $D_{\mathrm{sub}}$-dimensional sub-vectors, each independently quantized.
As a result, each data point is stored as a quantization code, eliminating the need to store the original high-dimensional vector. 
This approach significantly reduces the memory footprint of the \ac{VDB} and enables fast distance computation through the use of lookup tables.
During an \ac{ANN} search, Faiss first performs a nearest-neighbor search among the centroids and selects the $P_{\mathrm{IVF}}$ clusters, including the cluster whose centroid has the highest similarity to the query vector and its neighboring clusters.
A similarity search is subsequently conducted only within the selected clusters to retrieve the approximate nearest neighbor.
%
%
This approach avoids an exhaustive search across all data points and significantly reduces the search space by focusing only on data points that are likely to be close to the query vector.
The parameter $P_{\mathrm{IVF}}$ governs the trade-off between search accuracy and computational complexity, enabling efficient retrieval with low memory usage even for large-scale \ac{VDB}.

The cost of the \ac{VSS} is evaluated in terms of the number of search operations $N_\mathrm{s}$.
In an exhaustive nearest-neighbor search, referred to as the Flat index in the Faiss library, the number of search operations $N_\mathrm{s}$ for a single query vector is equal to the \ac{VDB} size and thus scales on the order of $N_{\mathrm{s}}$.
In contrast, for the IVFPQ index, which combines an \ac{IVF} index with \ac{PQ}, $N_\mathrm{s}$ grows on the order of
\begin{equation}
    \mathcal{O}\left(K_{\mathrm{IVF}} + P_{\mathrm{IVF}} \times \frac{D_{\mathrm{VDB}}}{K_{\mathrm{IVF}}}\right).
    \label{eq: search num.}
\end{equation}
The first term in \eqref{eq: search num.} represents the number of distance computations required for the nearest-centroid search, whereas the second term corresponds to the number of distance computations for data points within the selected clusters.
From \eqref{eq: search num.}, $N_\mathrm{s}$ is minimized when $K_{\mathrm{IVF}} = \sqrt{P_{\mathrm{IVF}} D_{\mathrm{VDB}}}$.
In this study, based on the guidelines provided in the official documentation~\cite{Douze2025} as well as empirical experiments from practical use, we adopt $K_{\mathrm{IVF}} = 4\sqrt{D_{\mathrm{VDB}}}$ as a setting that achieves a favorable balance between computational efficiency and search accuracy.
By appropriately adjusting $K_{\mathrm{IVF}}$, the search complexity $N_\mathrm{s}$ can be controlled to scale at most as $\mathcal{O} \left( \sqrt{D_{\mathrm{VDB}}} \right)$, thereby ensuring both efficiency and scalability.

Although it is possible to directly use the value returned by the \ac{ANN} search with the key $\acute{\bm{\upsilon}}_m$ generated in \eqref{eq:find_nu} as the predicted \ac{MI}, the inherent approximation in the \ac{ANN} search does not guarantee retrieval of the exact nearest neighbor.
Moreover, depending on the design of the feature vectors and the scale of the \ac{VDB}, the value associated with the nearest key may not necessarily yield the optimal \ac{MI} prediction.
To improve the robustness of the prediction, we also employ the Top-$K$ method for the \ac{ANN} output~\cite{Yang2023}.
Specifically, as illustrated in Fig.~\ref{fig:MIpre_VDB} (b), during the \ac{ANN} search, we extract the values associated with the top $K$ keys that yield the highest approximate vector similarity computed internally, and then use their average as the final \ac{MI} prediction, as given by
\begin{equation}
    \tilde{I}_{m,\mathrm{TB}}^Q = \frac{1}{K} \sum_{k=1}^{K} \varepsilon_k,
    \label{eq:top-K_mean}
\end{equation}
where $\varepsilon_1\ldots,\varepsilon_K$ are the values associated with the top $K$ keys.
The pseudocode of the \ac{VSS}-based \ac{MI} prediction using the Top-$K$ method is presented in Algorithm \ref{alg:Top-K}.


\begin{algorithm}[t]
\caption{VSS-based MI prediction using Top-$K$ method}
\label{alg:Top-K}
\hrulefill
\begin{algorithmic}[1]
\Require{Estimated CSI $\acute{\bm{H}}$, average SNR $\acute{\rho}$, \par \hspace{1.5ex} VDB for $Q$-QAM}
\Ensure{Predicted MI $\tilde{I}_{m,\mathrm{TB}}^Q$}
\vspace{-1ex}
\Statex \hspace{-4ex}\hrulefill
%
\State Compute the feature vector (key): $\acute{\bm{\upsilon}} = f(\acute{\bm{H}}, \acute{\rho})$
\State Execute Top-$K$ \ac{ANN} search using query $\acute{\bm{\upsilon}}$
\State Obtain top $K$ pairs in descending order of similarity: 
$(\bm{\upsilon}_1, \varepsilon_1), (\bm{\upsilon}_2, \varepsilon_2), \ldots, (\bm{\upsilon}_K, \varepsilon_K)$
\State Compute the average: 
$\tilde{I}_{m,\mathrm{TB}}^Q = \frac{1}{K} \sum_{k=1}^{K} \varepsilon_k$
\end{algorithmic}
\end{algorithm}


\subsection{Design of Feature Vector}
\label{subsec:FeatureVec}

A feature vector is a low-dimensional representation that robustly and sufficiently extracts the information required for the target task from an extremely high-dimensional parameter space characterizing the wireless environment.
In other words, its design is the key factor that determines the success or failure of the proposed approach.
Meanwhile, although wireless channels vary according to stochastic events, their variations exhibit strong correlations across the time, frequency, and spatial domains.
By designing a feature vector that effectively captures these characteristics, it becomes possible to achieve accurate predictions with a much smaller database size than that required for other applications, such as natural language processing and high-dimensional image classification.

%
%

The fundamental principle of feature vector design in this study is that, within the limits of practical feasibility, one should identify a low-dimensional representation that can analytically explain the target task, relying as much as possible on a mathematical approach.
In Section~\ref{sec: MI-based MCS Selection}, we demonstrated that the demodulator output \ac{MI} after \ac{LMMSE} detection can be analytically estimated on a \ac{TB}-wise basis.
On the other hand, analytically predicting the convergence behavior of \ac{EP} is intrinsically difficult.
Therefore, the demodulator output \ac{MI} after \ac{LMMSE} detection, which corresponds to the first iteration of \ac{EP}, is positioned as a limiting metric that can analytically explain the target task of predicting the demodulator output \ac{MI} after \ac{EP} detection.

Based on the above discussion, we construct the feature vector in this paper using the predicted \ac{MI} obtained with \ac{LMMSE}-based \ac{MUD} for each subcarrier, \textit{i.e.}, $\tilde{I}_{m}^Q[\ell]$ in \eqref{eq:PreMI_LMMSE}.
Considering all $L = 192$ subcarriers that constitute one \ac{TB}, the feature vector is given by\footnote{In fact, through a preliminary ablation study, we have confirmed that this mathematically grounded feature vector achieves clearly superior performance compared with other conceivable alternative designs.}
%
%
\begin{equation}
    \bm{\upsilon}_m = \left[\tilde{I}_{m}^Q[1],
    \tilde{I}_{m}^Q[2],\ldots, \tilde{I}_{m}^Q[192]
    \right]^{\mathsf{T}}\in\mathbb{R}^{192\times 1},
    \label{eq:featureVec_LMMSEMI}
\end{equation}
where the \ac{ANN} search over the \ac{VDB} can be interpreted as transforming the per-subcarrier \ac{MI} obtained with \ac{LMMSE}-based \ac{MUD} into the \ac{TB}-wise \ac{MI} with \ac{EP}-based \ac{MUD}.


\subsection{Evaluation of MI Prediction Accuracy}
\label{subsec:VSS_MI_res}

Computer simulations were conducted to evaluate the prediction accuracy achieved with the feature vector defined in \eqref{eq:featureVec_LMMSEMI}.
The simulation conditions were identical to those in Subsection \ref{subsec:EvaluationBLERPrediction}, \textit{i.e.}, the parameters listed in Tab. \ref{table: prm MIMO-OFDM}.
The \ac{VDB} size was fixed at $D_\mathrm{VDB}$ for all modulation schemes, and the data points ($220800$), corresponding to $200$ channels, were used for accuracy evaluation.
The channel realizations for evaluation were generated independently from those used for offline \ac{VDB} construction.
The average received \ac{SNR} was varied from $-10$ dB to $24$ dB in $0.5$ dB increments.
Consequently, the \ac{VDB} size is given by the product of the number of channel realizations, the number of \ac{SNR} points, and the number of \acp{UE}.
The internal \ac{ANN} search parameters were set to $(K_{\mathrm{IVF}},D_\mathrm{sub},P_{\mathrm{IVF}}) = (4\sqrt{D_{\mathrm{VDB}}},12,10)$,

First, to demonstrate the practical feasibility of the proposed method, we evaluate the online search latency of \ac{VSS} implemented on a \ac{GPU} using the Faiss library.
Although the processing latency in practical deployments depends on the system configuration and execution conditions, all simulation results presented in this paper were obtained on a commercially available personal computer equipped with an NVIDIA GeForce RTX 4070 SUPER GPU.
%

\begin{figure}[t]
\centering
\includegraphics[width=0.9\columnwidth,keepaspectratio=true]{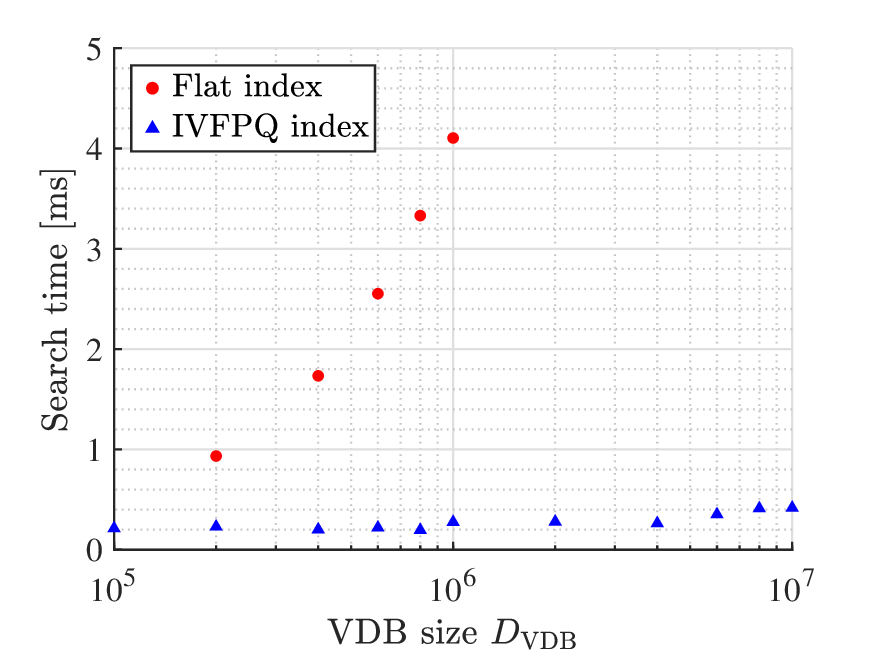}
\vspace{-2mm}
\caption{Search time versus \ac{VDB} size under a worst-case evaluation scenario, where the \ac{VDB} stores $192$-dimensional real-valued feature vectors with independently and uniformly distributed elements in the range $[0, 1]$.}
\label{fig: Search Time}
\vspace{-4mm}
\end{figure}

Fig.~\ref{fig: Search Time} shows the search time as a function of the \ac{VDB} size $D_{\mathrm{VDB}}$ for each index type, namely, the Flat index and the IVFPQ index.
The horizontal axis represents $D_{\mathrm{VDB}}$, while the vertical axis shows the search time per query vector.
For the Flat index, since distance computations are required for all data points, the search time increases rapidly in proportion to the \ac{VDB} size.
In contrast, for the IVFPQ index (\textit{i.e.}, \ac{ANN} search), the search time remains almost constant even as the \ac{VDB} size increases.
This behavior arises because the search is restricted to a small number of clusters in the vicinity of the query vector, indicating that an increase in the \ac{VDB} size has little impact on the actual number of candidate points examined during the search.
Although the required \ac{VDB} size is discussed later, these results indicate that, with fast \ac{ANN}-based \ac{VSS}, the search time can be reduced to the order of several hundred microseconds even on a commercially available \ac{GPU}.
Moreover, owing to its high scalability, we conclude that the \ac{VSS} search time does not constitute a dominant computational bottleneck in the proposed method.

In the following, we first evaluate the variation in prediction accuracy and robustness with respect to the \ac{VDB} size and the Top-$K$ parameter.
Based on these results, we then assess the \ac{MI} prediction accuracy of the proposed method using the selected parameter settings.

\subsubsection{Impact of the VDB Size}

\begin{figure*}[t]
\begin{center}
    \subfloat[$4$QAM.]{
    \includegraphics[width=0.66\columnwidth,keepaspectratio=true]{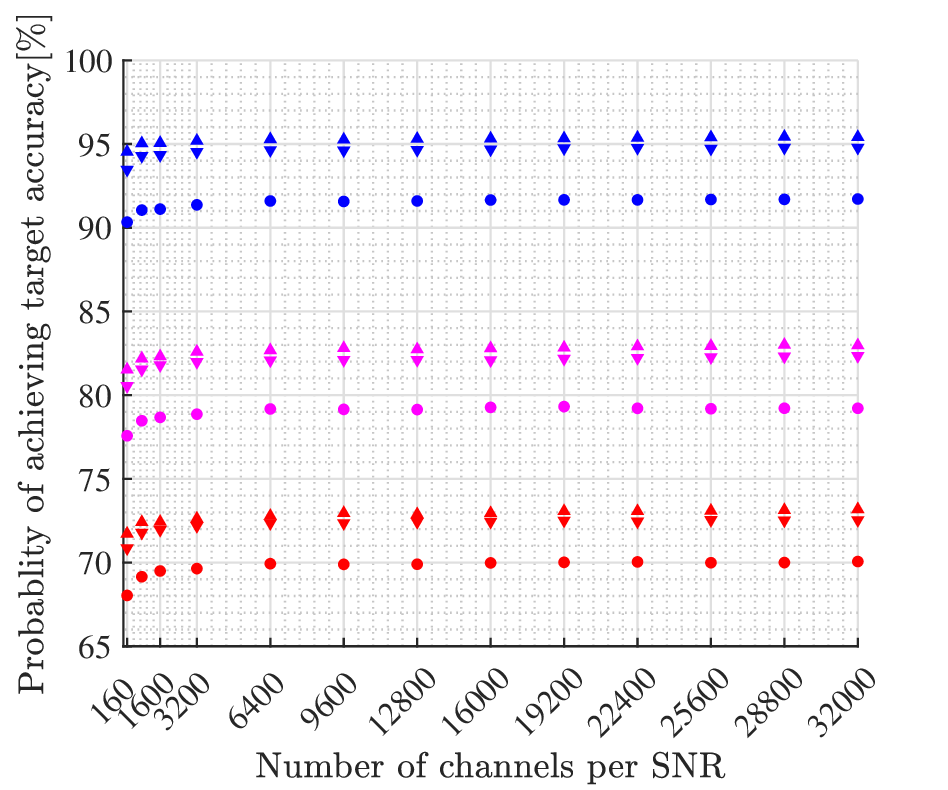}
    }
    \subfloat[$16$QAM.]{
    \includegraphics[width=0.66\columnwidth,keepaspectratio=true]{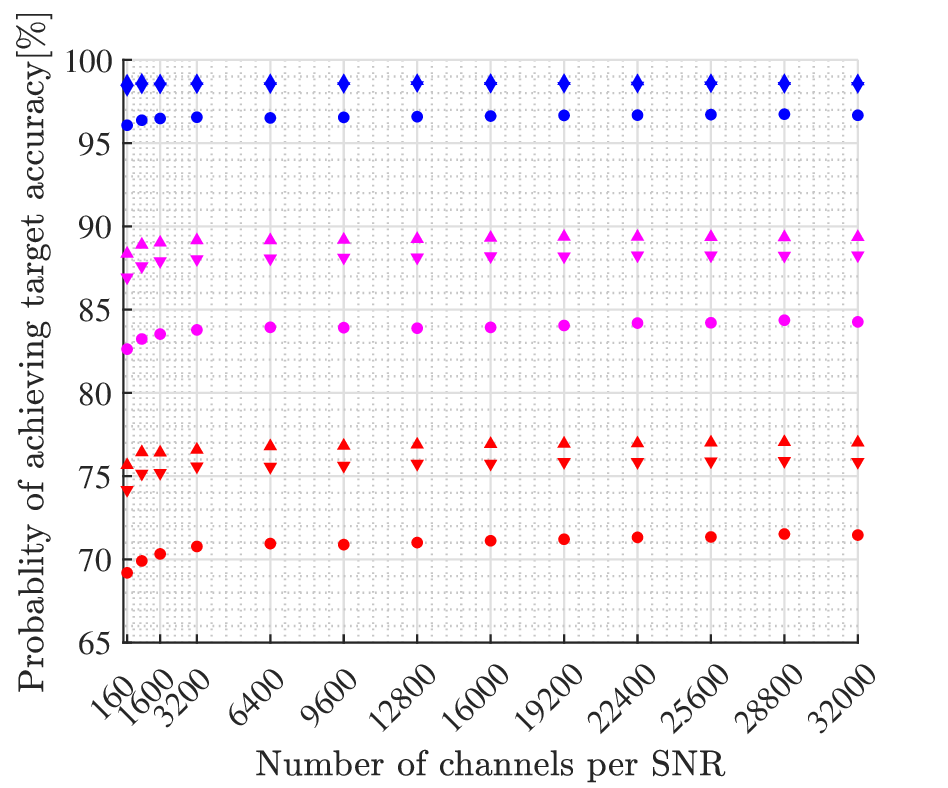}
    }
    \subfloat[$64$QAM.]{
    \includegraphics[width=0.66\columnwidth,keepaspectratio=true]{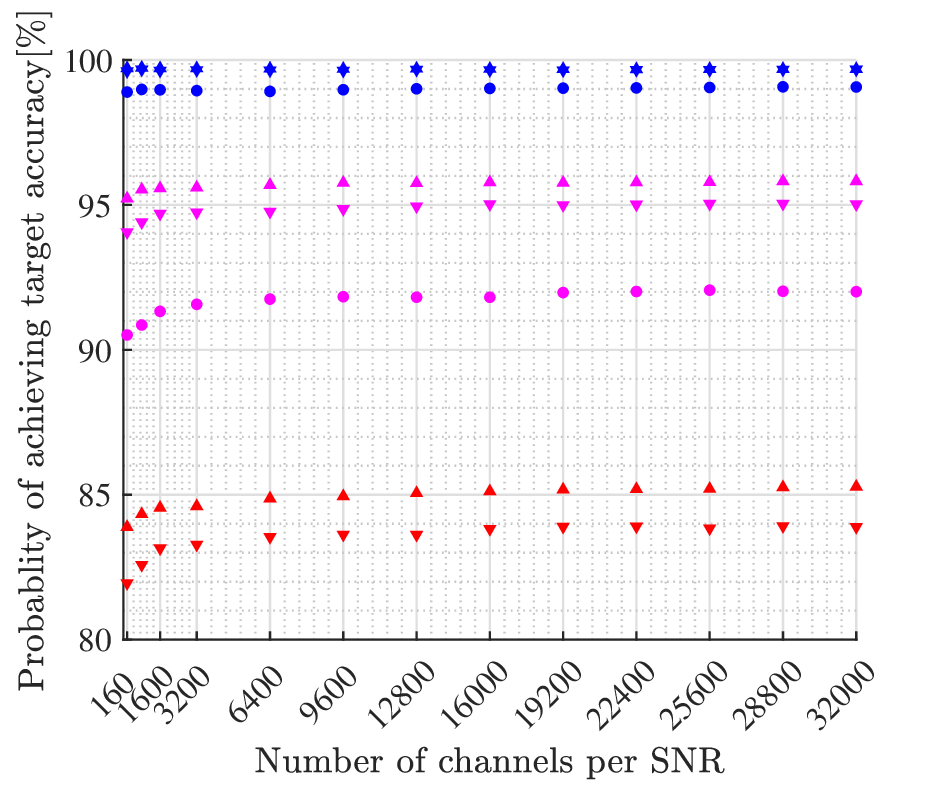}
    }
    \\
    \vspace{-2mm}
    \subfloat{
    \includegraphics[width=1.1\columnwidth,keepaspectratio=true]{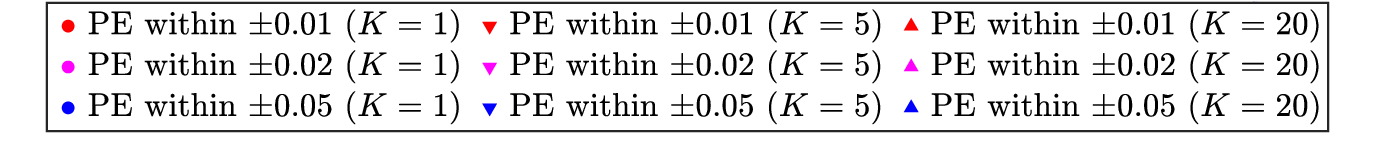}
    }
    \vspace{-2mm}
    \caption{Prediction performance versus \ac{VDB} size, evaluated by the probability that the \ac{MI} prediction error falls within specified bounds.}
    \label{fig: Accuracy (VDB size)}
\end{center}
\vspace{-8mm}
\end{figure*}

\begin{figure*}[t]
\begin{center}
    \subfloat[$4$QAM.]{
    \includegraphics[width=0.66\columnwidth,keepaspectratio=true]{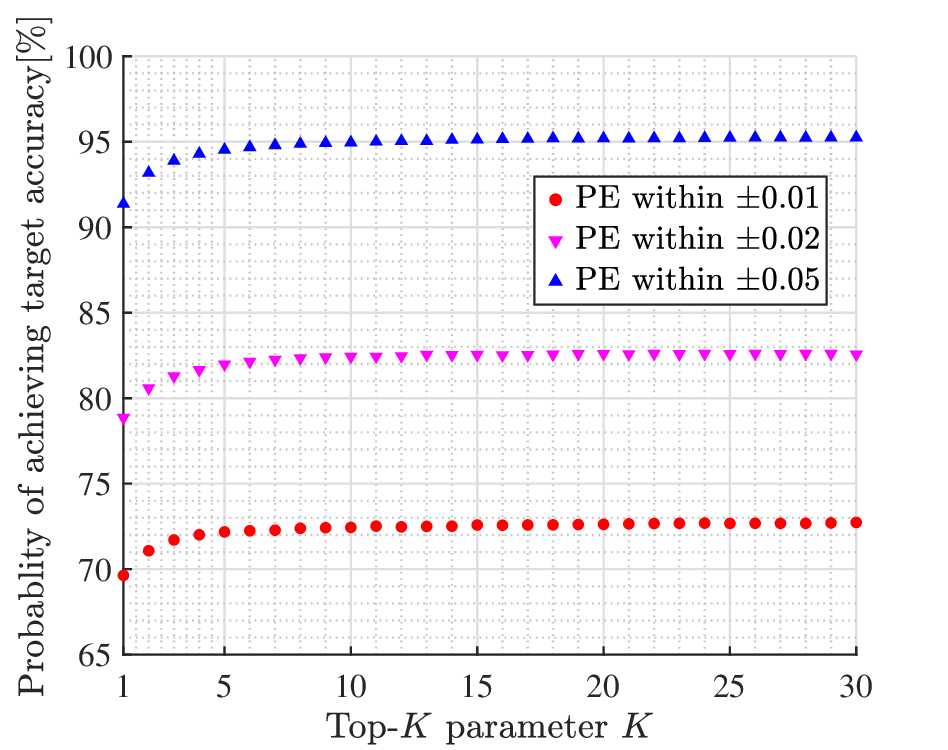}
    }
    \subfloat[$16$QAM.]{
    \includegraphics[width=0.66\columnwidth,keepaspectratio=true]{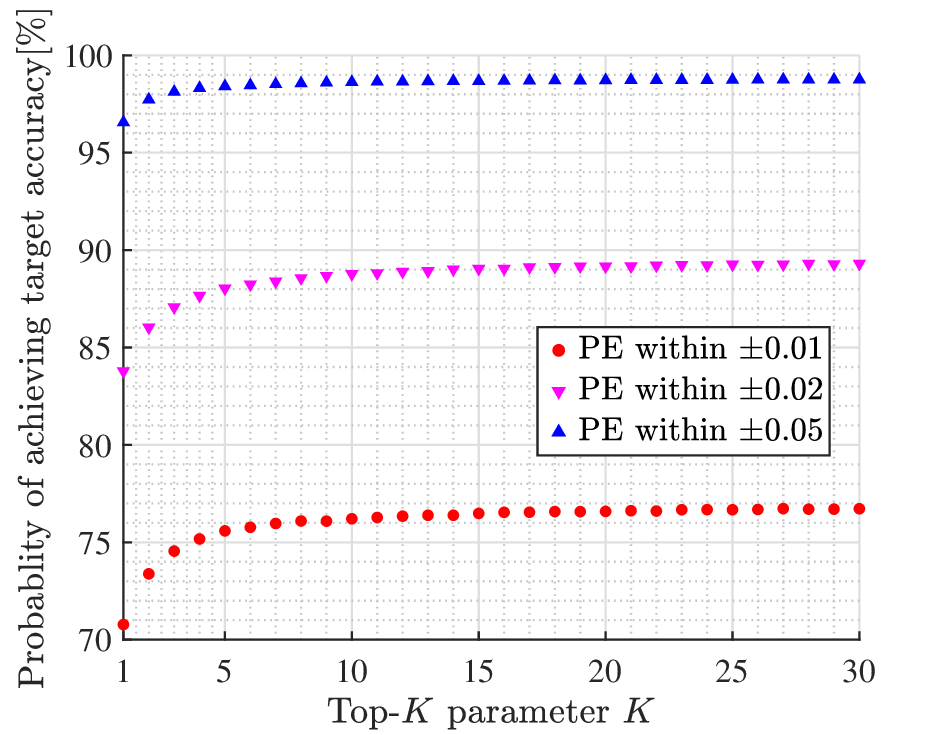}
    }
    \subfloat[$64$QAM.]{
    \includegraphics[width=0.66\columnwidth,keepaspectratio=true]{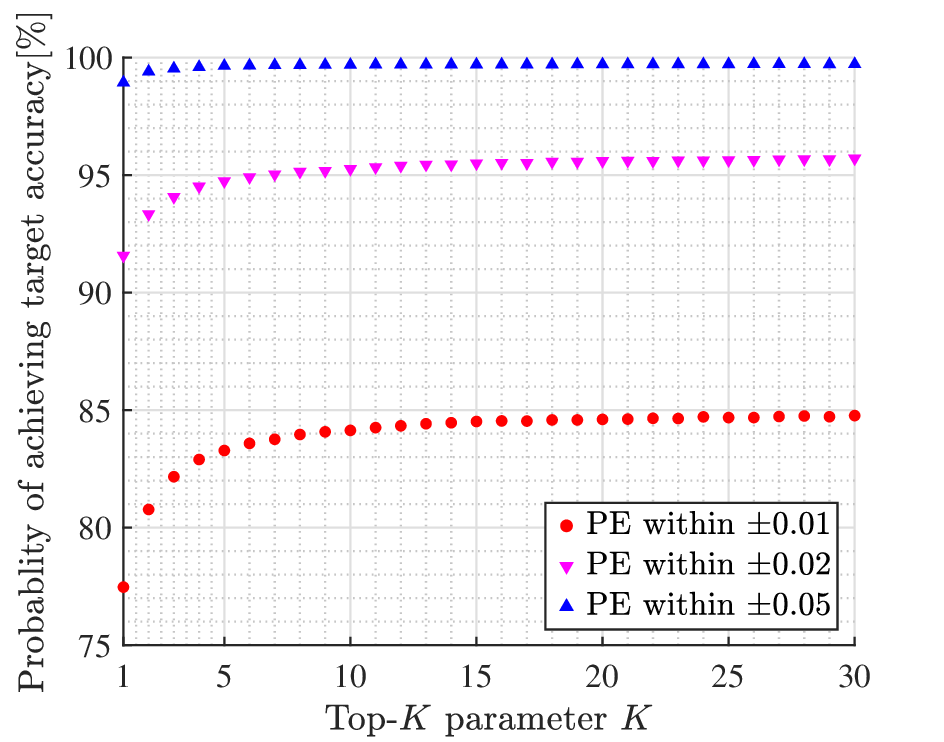}
    }
    \caption{Prediction performance versus Top-$K$ parameter, evaluated by the probability that the \ac{MI} prediction error falls within specified bounds.}
    \label{fig: Accuracy (parameter:K)}
\end{center}
\vspace{-6mm}
\end{figure*}

Fig.~\ref{fig: Accuracy (VDB size)} illustrates the variation in prediction accuracy as a function of the \ac{VDB} scale.
The horizontal axis represents the number of user channel realizations for each average received \ac{SNR}, which determines the database scale and is proportional to the \ac{VDB} size.
The vertical axis shows the probability that the prediction error, defined as $\tilde{I}_{m, \mathrm{TB}}^{Q} - I_{m, \mathrm{TB}}$, falls within the ranges of $\pm 0.01$, $\pm 0.02$, and $\pm 0.05$.
The Top-$K$ parameter is set to $K = 1$, $5$, and $20$, respectively.

From the results, it can be observed that, for all modulation schemes, once the number of user channel realizations exceeds approximately $3,200$ (\textit{i.e.}, $200$ channel realizations $\times$ $16$ \acp{UE}), the prediction accuracy becomes almost insensitive to further increases in the \ac{VDB} scale.
This indicates that approximately $3,200$ user channel realizations are sufficient to capture the statistical characteristics of the \ac{CDL}-B channel model compliant with the \ac{5G NR} specification that are required to accomplish the target task of predicting the demodulator output \ac{MI} after \ac{EP} detection.
At the same time, this observation suggests that the effectiveness of the proposed approach strongly depends on the appropriately designed feature vector.
Furthermore, these findings not only indicate that offline \ac{VDB} construction does not incur excessive computational costs but also suggest the feasibility of online \ac{VDB} updates.
%
%

\subsubsection{Impact of the Top-$K$ Parameter}

Next, based on the results of the previous experiments, we evaluate the impact of the Top-$K$ parameter on the prediction accuracy by fixing the number of user channel realizations per average received \ac{SNR} to $3,200$.
Fig.~\ref{fig: Accuracy (parameter:K)} illustrates the variation in prediction accuracy as a function of the Top-$K$ parameter.
The horizontal axis represents the number of samples $K$ used for averaging in the Top-$K$ method, while the vertical axis shows, as in Fig.~\ref{fig: Accuracy (VDB size)}, the probability that the prediction error falls within a prescribed range.

From Fig.~\ref{fig: Accuracy (parameter:K)}, it can be observed that, for all modulation schemes, the prediction accuracy improves monotonically as the averaging parameter $K$ increases.
This is because the candidates returned by the \ac{ANN} search are approximate rather than exact nearest neighbors.
When only a single search result is used, the approximation error inherent in \ac{ANN} search can degrade prediction accuracy.
In contrast, averaging the prediction values obtained from multiple \ac{ANN} outputs effectively generalizes the estimation and mitigates the impact of approximation errors.
%
Although increasing $K$ enlarges the scale of parallel searches, values of $K$ on the order of several tens—sufficient to achieve adequate generalization performance—do not pose practical computational burdens even on commercially available \acp{GPU}.
Furthermore, as expected, the improvement in prediction accuracy exhibits diminishing returns as $K$ becomes sufficiently large and eventually saturates.
Under the experimental conditions considered in this study, we confirmed that $K=20$ provides a favorable trade-off between prediction accuracy and the degree of parallel search.

\subsubsection{MI Prediction Accuracy}

Based on the above results, we set the parameters accordingly and evaluate the \ac{MI} prediction accuracy.
A total of $400$ \acs{MU-MIMO-OFDM} channel realizations satisfying the parameter settings in Tab.~\ref{table: prm MIMO-OFDM} were generated, of which $200$ were used for \ac{VDB} construction and the remaining $200$ for evaluation.
In this setup, the \ac{VDB} size is given by the product of the number of channel realizations, the number of \acp{UE}, and the number of \ac{SNR} points, \textit{i.e.}, $D_{\mathrm{VDB}} = 200 \times 16 \times 69 = 220800$.
The internal \ac{ANN} search parameters are set to $(K_{\mathrm{IVF}},D_\mathrm{sub},P_{\mathrm{IVF}}) = (1879,12,10)$, and the Top-$K$ parameter is set to $K = 20$.
%

\begin{table}[!t]
  \vspace{-6mm}
  \caption{Prediction accuracy}
  \label{table:res}
  \vspace{-2mm}
  \centering
  \scalebox{0.90}{
  \begin{tabular}{|c|c|c|c|} 
    \hline
    {\cellcolor[rgb]{0.95, 0.95, 0.95}} Modulation scheme & 
    {\cellcolor[rgb]{0.95, 0.95, 0.95}} $4$QAM & 
    {\cellcolor[rgb]{0.95, 0.95, 0.95}} $16$QAM &
    {\cellcolor[rgb]{0.95, 0.95, 0.95}} $64$QAM \\
    \hline \hline
    PE within $\pm0.01$ & $72.61 \%$ & $76.57 \%$ & $84.67 \%$ \\ \hline
    PE within $\pm0.02$ & $82.60 \%$ & $89.13 \%$ & $95.64 \%$ \\ \hline
    PE within $\pm0.05$ & $95.21 \%$ & $98.73 \%$ & $99.70 \%$ \\ \hline \hline
    PE over $+0.01$ & $12.23 \%$ & $11.17 \%$ & $7.66 \%$ \\ \hline
    PE over $+0.02$ & $7.47 \%$ & $4.67 \%$ & $1.71 \%$ \\ \hline
    PE over $+0.05$ & $1.27 \%$ & $0.12 \%$ & $0.014 \%$ \\ 
    \hline \hline
    MSE & $0.000462$ & $0.000185$ & $0.0000676$ \\
    \hline
  \end{tabular}
  }
  \vspace{-6mm}
\end{table}

\begin{figure*}[!t]
\begin{center}
    \subfloat[EP detection, $4$QAM.]{
    \includegraphics[width=0.66\columnwidth,keepaspectratio=true]{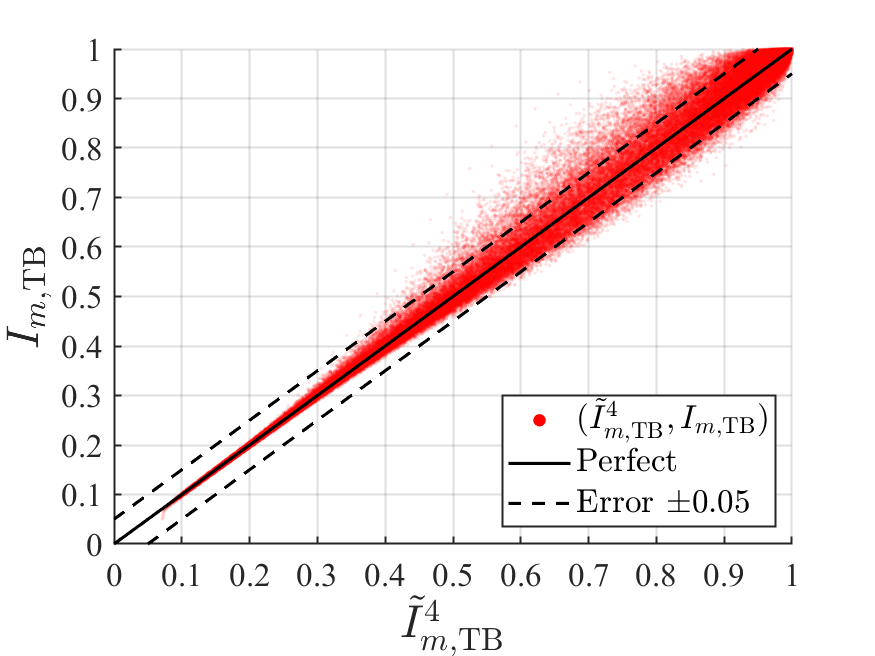}
    }
    \subfloat[EP detection, $16$QAM.]{
    \includegraphics[width=0.66\columnwidth,keepaspectratio=true]{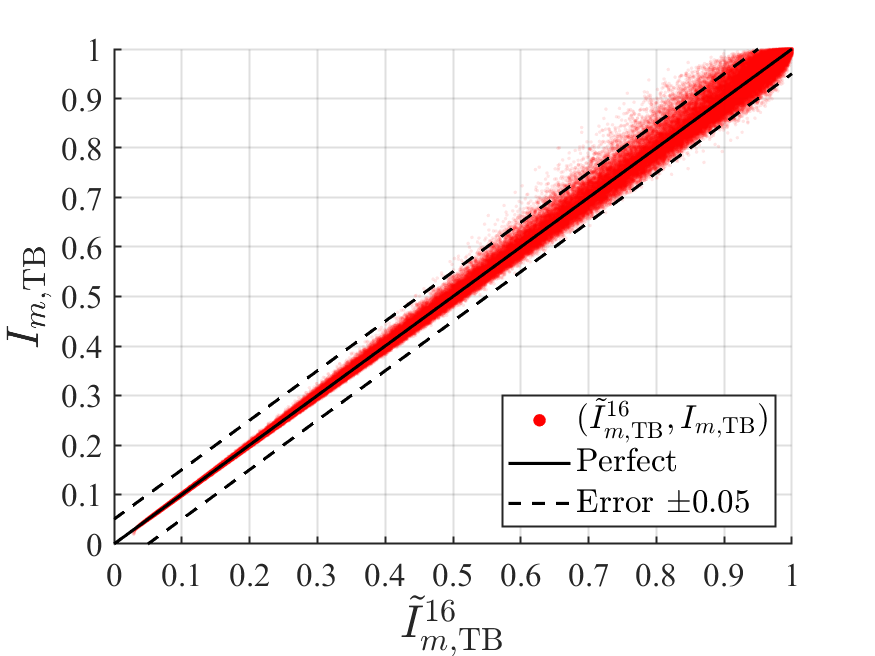}
    }
    \subfloat[EP detection, $64$QAM.]{
    \includegraphics[width=0.66\columnwidth,keepaspectratio=true]{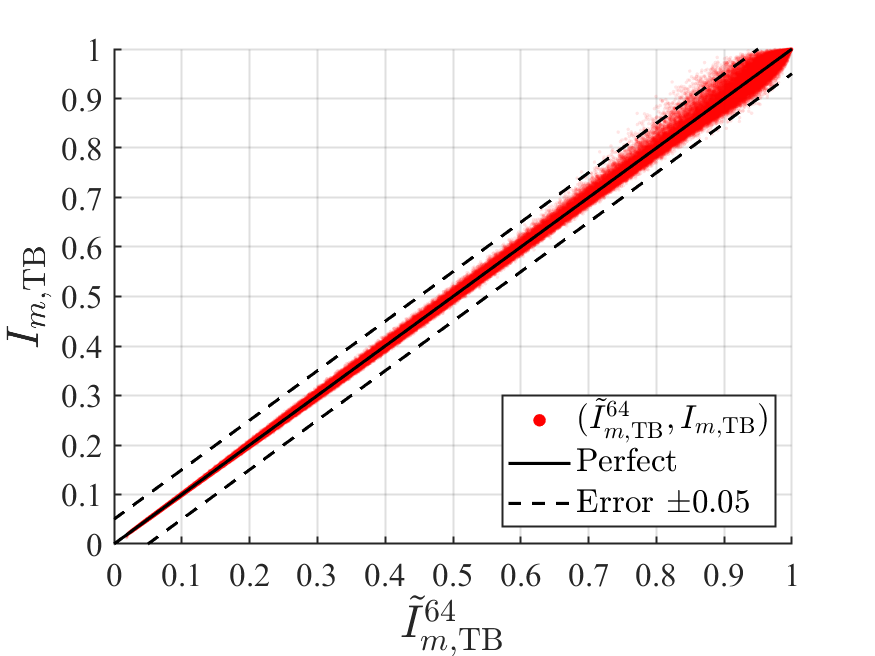}
    }
    \caption{Scatter plots of the measured \ac{MI} $I_{m,\mathrm{TB}}$ versus the predicted \ac{MI} $\tilde{I}_{m,\mathrm{TB}}^Q$ based on the proposed method.}
    \label{fig:res_VDB}
    \vspace{-4mm}
\end{center}
\end{figure*}

\begin{figure*}[t]
\centering
\includegraphics[width=2.0\columnwidth,keepaspectratio=true]{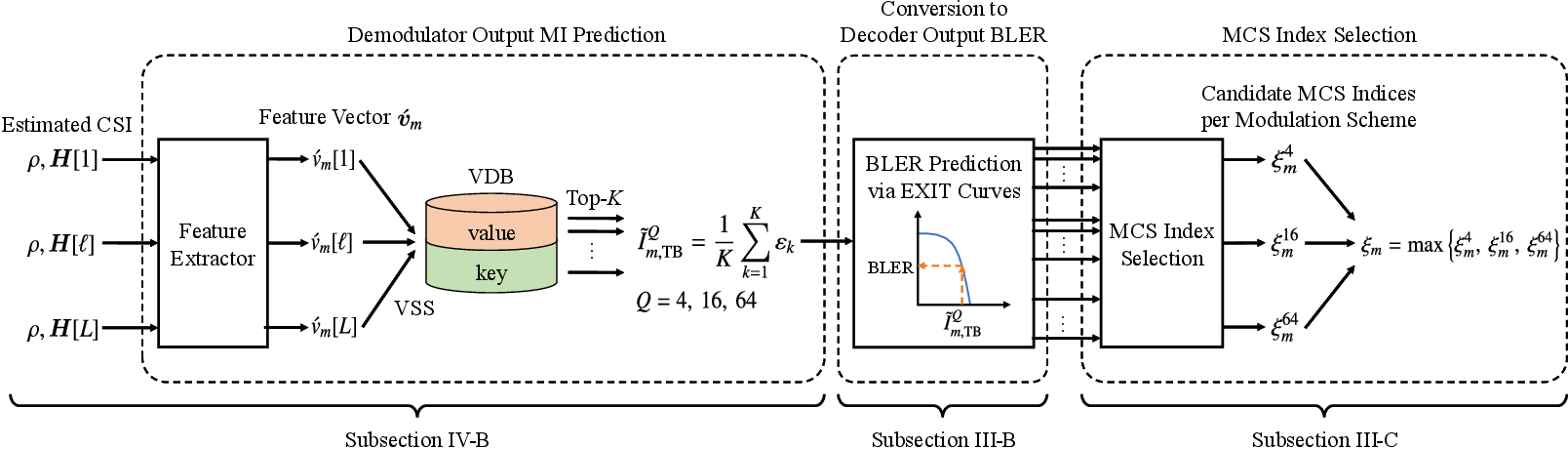}
\caption{Block diagram of the proposed \ac{MI}-based \ac{MCS} selection process incorporating the \ac{VSS}-based \ac{MI} prediction method.}
\label{fig: MI-based MCS selection proposed}
\vspace{-4mm}
\end{figure*}

Fig.~\ref{fig:res_VDB} shows scatter plots of the predicted and measured \ac{MI} pairs $(\tilde{I}_{m,\mathrm{TB}}^Q,I_{m,\mathrm{TB}})$ in the same manner as Fig. \ref{fig:res_Mean}.
Tab. \ref{table:res} summarizes the percentage of samples with a \ac{PE} within $\pm 0.01$, $\pm 0.02$, and $\pm 0.05$; the percentage exceeding the measured values by more than $+0.01$, $+0.02$, and $+0.05$; and the \acs{MSE} between the predicted and measured \acp{MI}. 
From Fig.~\ref{fig:res_VDB}, the sample points are concentrated near the ideal prediction line $I_{m,\mathrm{TB}} = \tilde{I}_{m,\mathrm{TB}}^Q$, and their variance decreases as the modulation order increases.
From the numerical results in Tab. \ref{table:res}, for all modulation schemes, more than $95\%$ of the samples have a \ac{PE} within $\pm0.05$, and more than $70\%$ are within $\pm0.01$.
Since a \ac{PE} exceeding $0.05$ could lead to selecting an adjacent \ac{MCS} index, this accuracy can be considered sufficient for the intended purpose.
Furthermore, the percentage of \ac{MI} overestimations relative to the measured values—which could increase the \ac{BLER}—is less than $2\%$ for a \ac{PE} threshold of $0.05$ ($\textit{i.e.}, \tilde{I}_{m,\mathrm{TB}}^Q \geq I_{m,\mathrm{TB}} + 0.05)$, and is even smaller for higher-order modulation schemes, where such errors are more critical.
These results demonstrate that the proposed method, using \eqref{eq:featureVec_LMMSEMI} as the key, can predict the post-\ac{MUD} \ac{MI} for \ac{EP}-based \ac{MUD} with sufficient accuracy to enable appropriate \ac{MCS} index selection.

\section{Throughput Evaluation}
\label{sec: Throughput Evaluation}

Finally, the effectiveness of the proposed \ac{MCS} selection framework is validated from a throughput perspective.

\vspace{-1mm}
\subsection{Proposed VSS-aided \ac{MCS} Selection Framework}

Fig.~\ref{fig: MI-based MCS selection proposed} illustrates the overall workflow of the \ac{MCS} selection process incorporating the proposed \ac{VSS}-based \ac{MI} prediction scheme, corresponding to Fig.~\ref{fig: MI-based MCS selection analytical} in Section \ref{sec: MI-based MCS Selection}.
Note that the demodulator output (post-\ac{MUD}) \ac{MI} prediction block in Fig.~\ref{fig: MI-based MCS selection analytical} is replaced with the proposed method described in Section IV.
The key advantage of the proposed framework is its applicability to \textit{any} \ac{MUD} scheme whose decoder input cannot be derived analytically.
As a representative example, this paper investigates \ac{EP}-based \ac{MUD}, well known for its high detection accuracy, and verifies whether its performance gains can be translated into throughput improvements.


\subsection{Throughput Evaluation}

\begin{figure*}[!t]
\begin{center}
    \subfloat[\ac{SNR} $=4$ dB.]{
    \includegraphics[width=0.66\columnwidth,keepaspectratio=true]{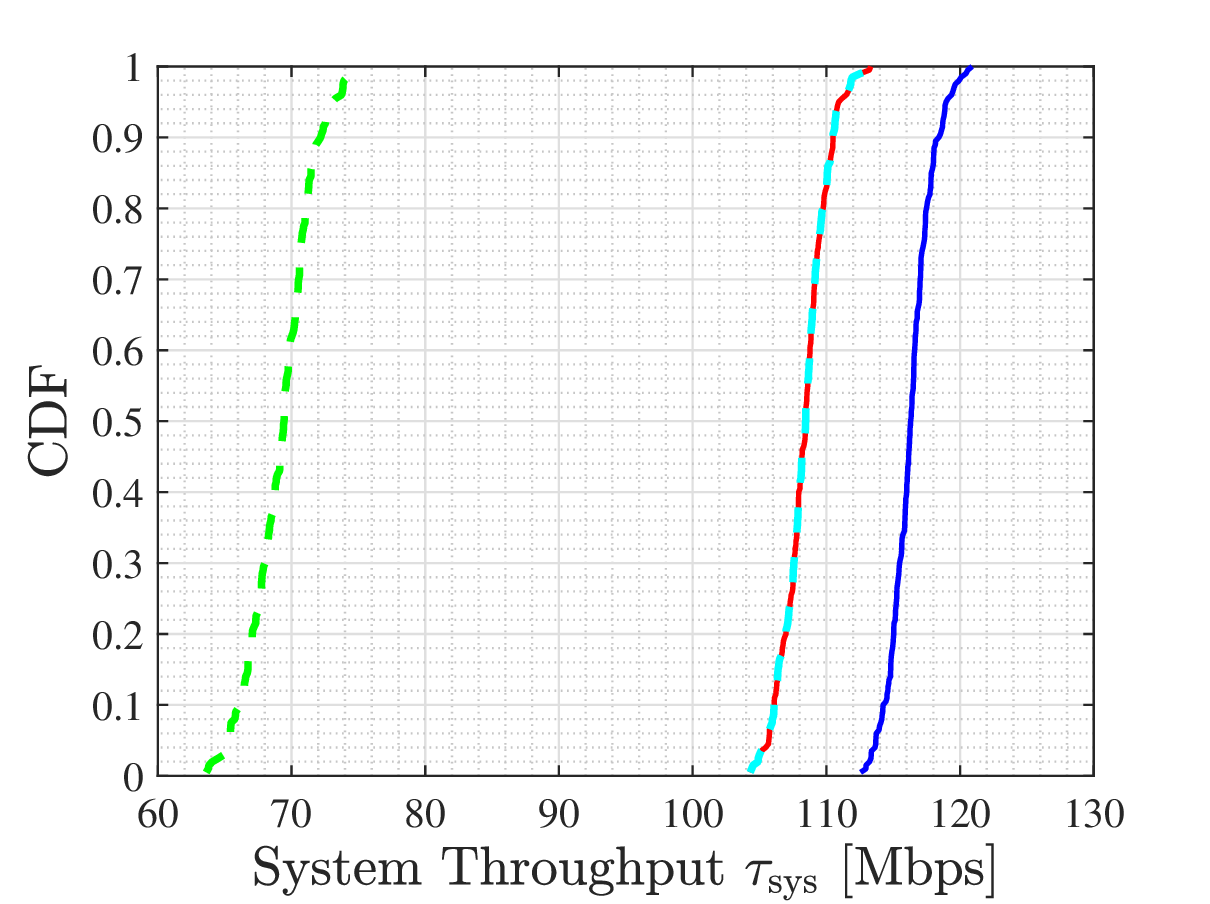}
    }
    \subfloat[\ac{SNR} $=12$ dB.]{
    \includegraphics[width=0.66\columnwidth,keepaspectratio=true]{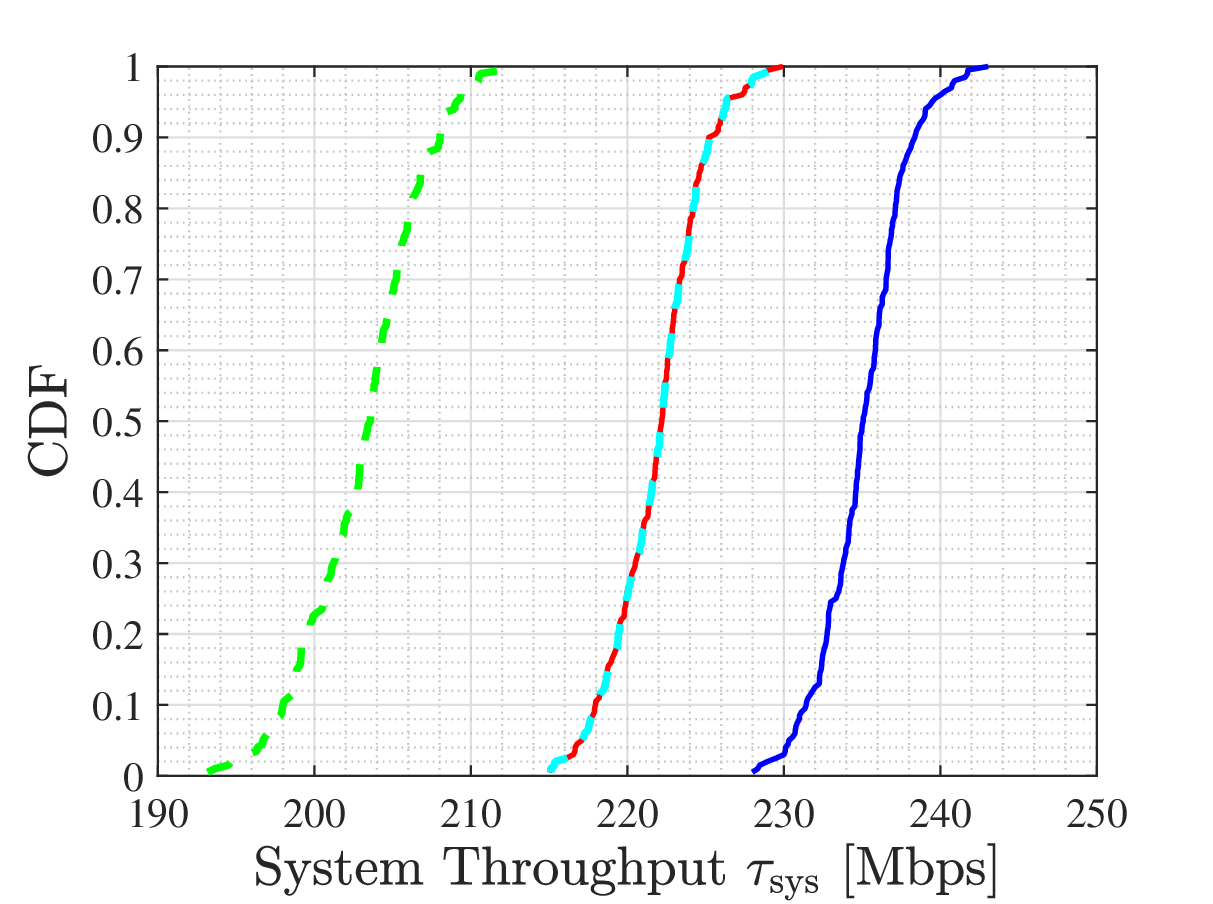}
    }
    \subfloat[\ac{SNR} $=20$ dB.]{
    \includegraphics[width=0.66\columnwidth,keepaspectratio=true]{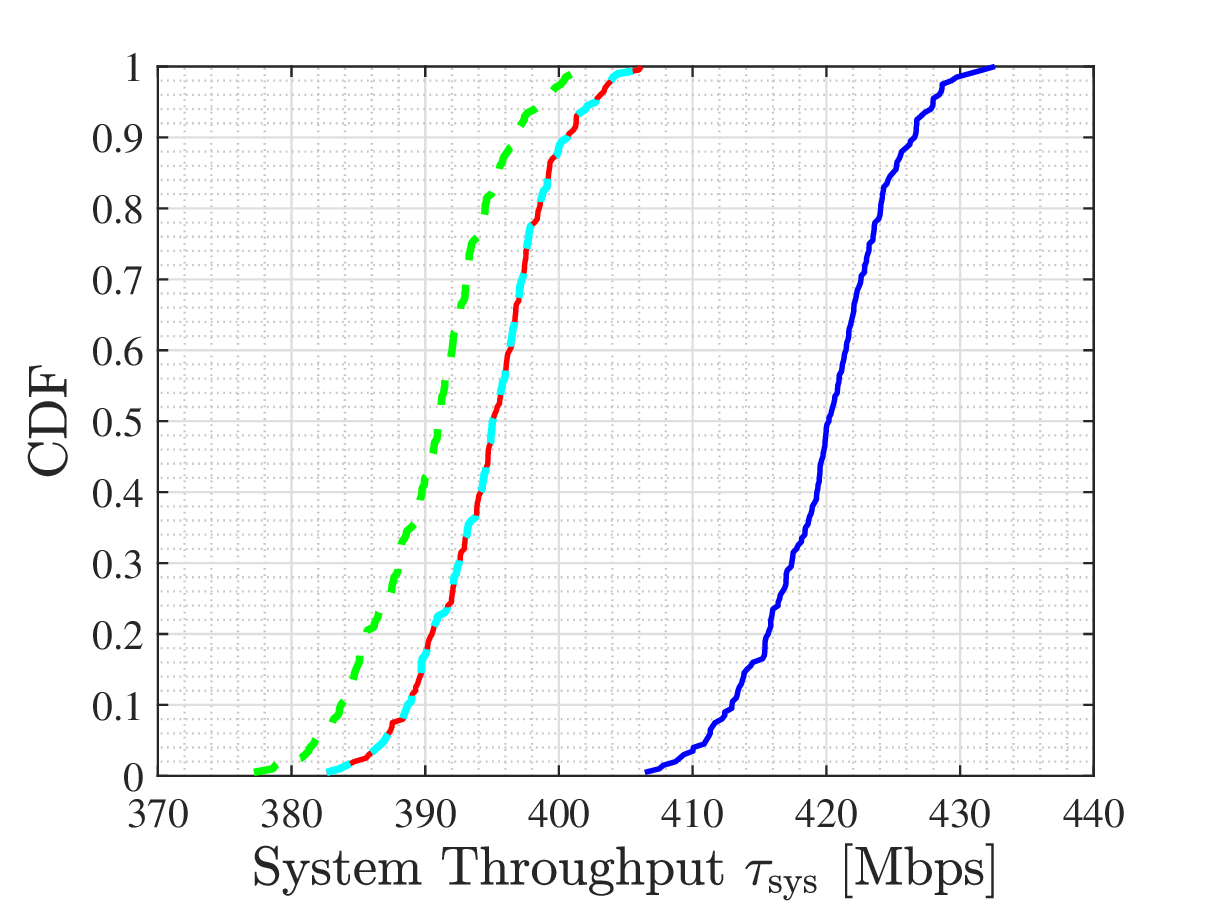}
    }
    \\
    \vspace{-2mm}
    \subfloat{
    \includegraphics[width=1.5\columnwidth,keepaspectratio=true]{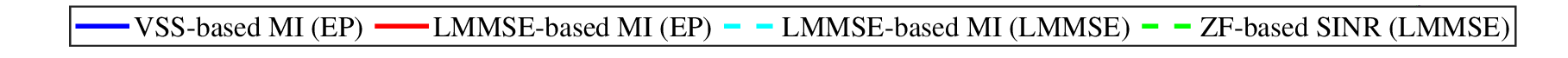}
    }
    \vspace{-2mm}
    
    \caption{System throughput performance for various average received \ac{SNR} levels in an \ac{MU-MIMO-OFDM} system.}
    \label{fig:sys_throughput}
\end{center}
\vspace{-8mm}
\end{figure*}

\begin{figure*}[!t]
\begin{center}
    \subfloat[\ac{SNR} $=4$ dB.]{
    \includegraphics[width=0.66\columnwidth,keepaspectratio=true]{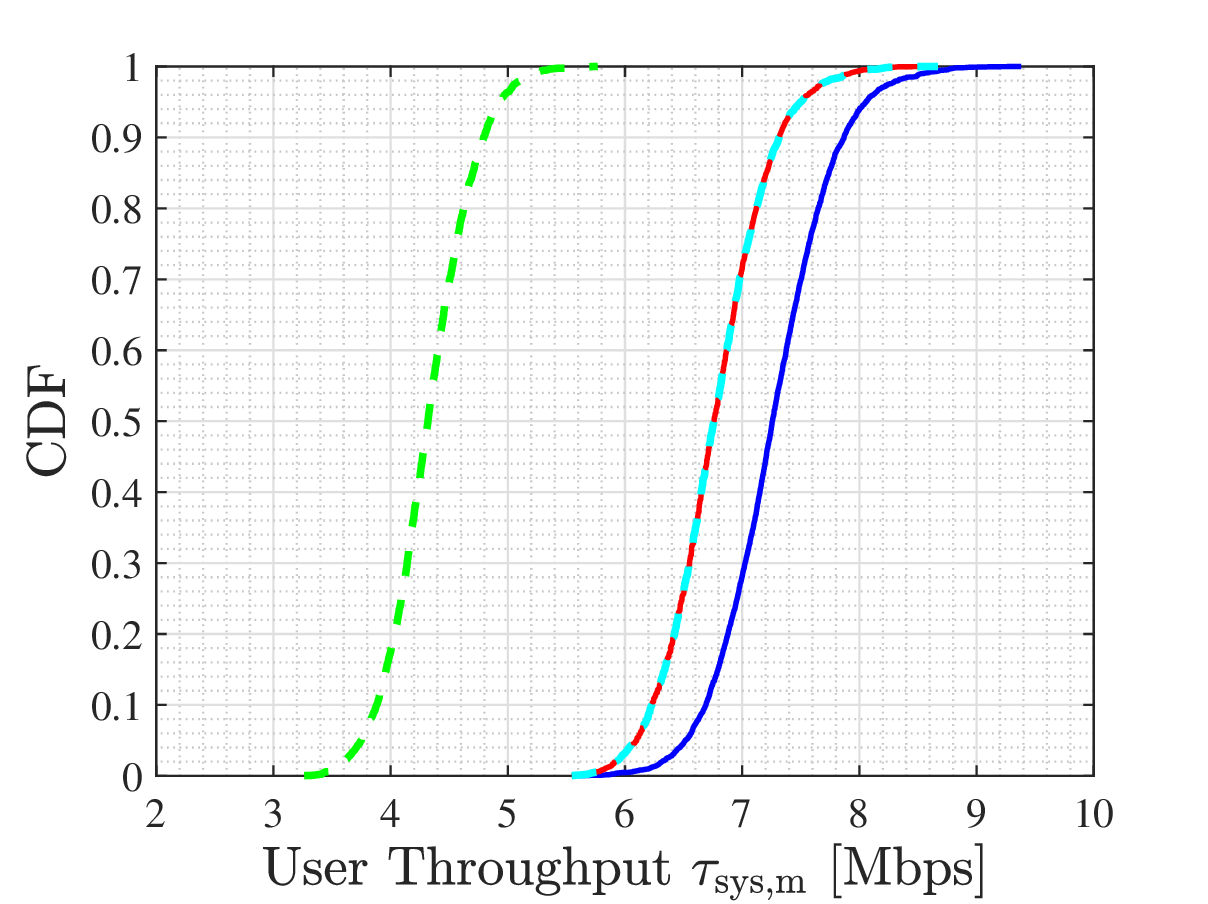}
    }
    \subfloat[\ac{SNR} $=12$ dB.]{
    \includegraphics[width=0.66\columnwidth,keepaspectratio=true]{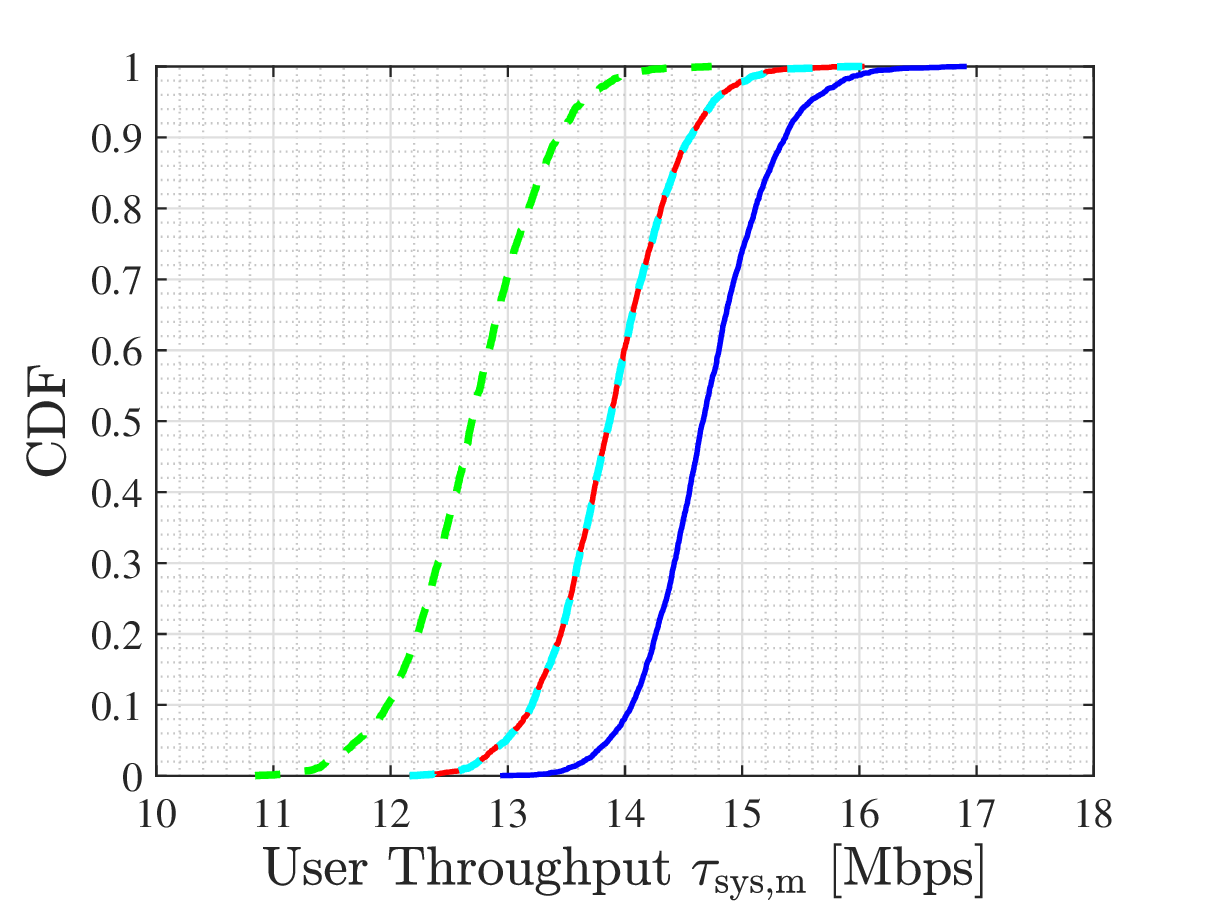}
    }
    \subfloat[\ac{SNR} $=20$ dB.]{
    \includegraphics[width=0.66\columnwidth,keepaspectratio=true]{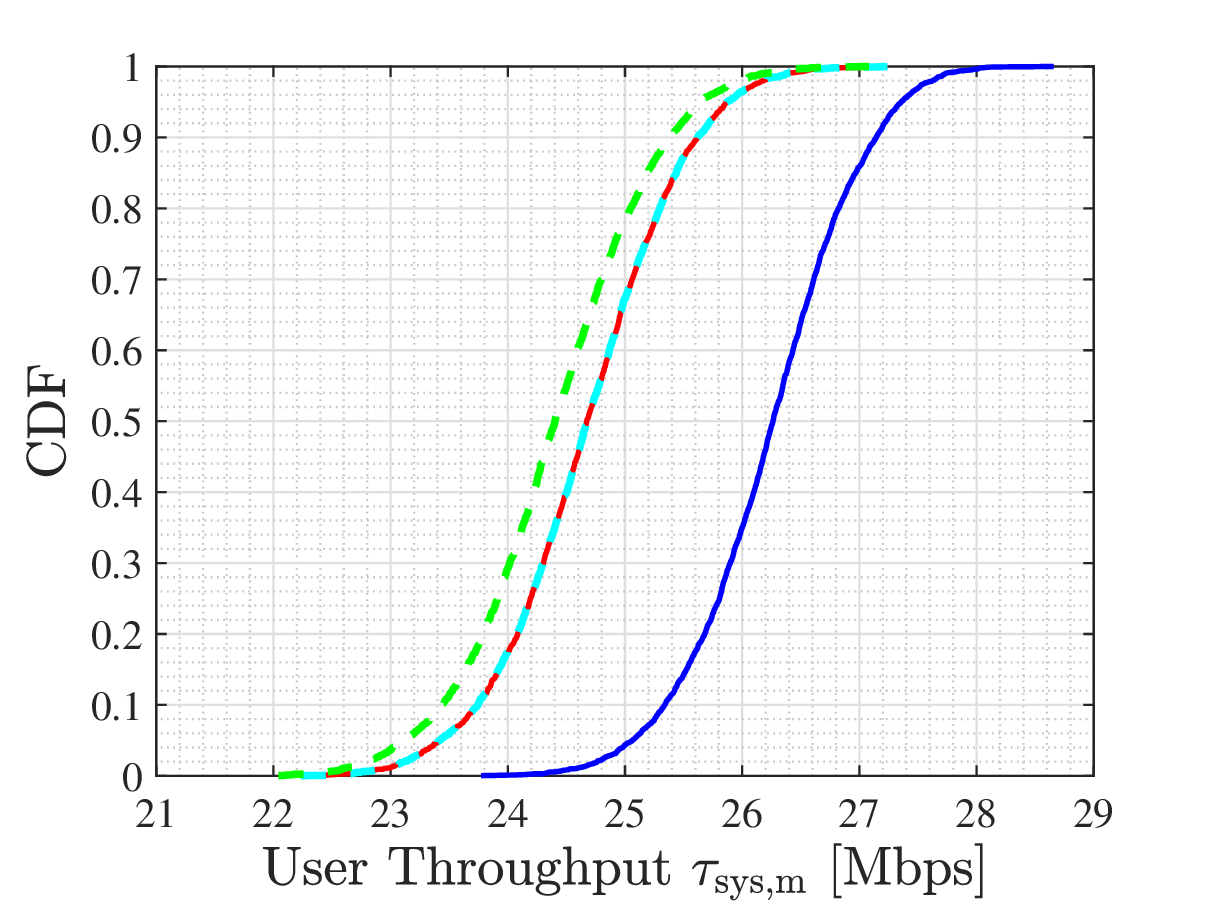}
    }
    \\
    \vspace{-2mm}
    \subfloat{
    \includegraphics[width=1.5\columnwidth,keepaspectratio=true]{FIG/Thoughput/legend.eps}
    }
    \vspace{-2mm}
    
    \caption{User throughput performance for various average received \ac{SNR} levels in an \ac{MU-MIMO-OFDM} system.}
    \label{fig:user_throughput}
\end{center}
\vspace{-6mm}
\end{figure*}

The simulation conditions are identical to those in Section \ref{sec: MI-based MCS Selection}, as listed in Tab.~\ref{table: prm MIMO-OFDM}, and the \ac{VDB} used for \ac{MI} prediction is the same as in Section \ref{sec: Proposed Method}.
The \ac{MCS} table follows the \ac{5G NR} standard~\cite{5GNR_MCS}, enabling \ac{BLER} prediction using the \ac{EXIT} curves shown in Fig. \ref{fig: EXIT curve} in Subsection \ref{subsec:ConversionMI}.
The reference \ac{BLER} for \ac{MCS} selection is set to $0.001$\footnote{
While a reference \ac{BLER} of $0.1$ is commonly adopted in practice~\cite{Durán2015,Peralta2022}, this study employs a more stringent value to ensure highly reliable transmission with advanced detectors. 
This also mitigates the risk of throughput degradation due to overestimation of channel quality.
}.
For system performance evaluation, $200$ independently generated channels are used.
These channels are identical to those used for evaluating \ac{MI} prediction accuracy in Subsection~\ref{subsec:VSS_MI_res}.
Throughput is calculated by performing \acs{OFDM} transmission over $N_{\mathrm{slot}}$ time slots.
%
Small-scale fading varies across slots, and \ac{MCS} selection is updated every $20$ slots,\footnote{
According to the \ac{5G NR} standard~\cite{5GNR_OFDM_RB}, when the subcarrier spacing is $30$ kHz, one frame consists of two subframes, each containing $10$ time slots.
} corresponding to one frame~\cite{5GNR_OFDM_RB}.
The user throughput is calculated as the sum of bits successfully transmitted without error per \ac{TB} in each time slot by each \ac{UE}, as follows:
\begin{equation}
    \tau_{\mathrm{user}, m}
    \triangleq
    \frac{\sum_{n_{\mathrm{slot}} = 1}^{N_{\mathrm{slot}}} N_{\mathrm{sucbit},m}(n_{\mathrm{slot}})}{N_{\mathrm{slot}} \cdot T_{\mathrm{slot}}} \:\:\: [\mathrm{Mbps}],
\end{equation}
where $N_{\mathrm{sucbit},m}(n_{\mathrm{slot}})$ [Mbit] is the number of information bits transmitted by the $m$-th \ac{UE} in time slot $n_{\mathrm{slot}}$ and successfully received, counted only when the entire \ac{TB} is error-free.
The slot duration $T_\mathrm{slot}$ is set to $0.5\times10^{-3}$ s, based on the \ac{5G NR} specification with a $30\,$kHz subcarrier spacing~\cite{5GNR_OFDM_RB}.
The system throughput is calculated from the total number of successfully transmitted bits across all \acp{UE} as follows:
\begin{eqnarray}
    \tau_{\mathrm{sys}} 
    \!\!\!\!&\triangleq&\!\!\!\! 
    \sum_{m = 1}^{M} \tau_{\mathrm{user}, m} \nonumber \\
    \!\!\!\!&=&\!\!\!\! 
    \frac{\sum_{m = 1}^{M} \sum_{n_{\mathrm{slot}} = 1}^{N_{\mathrm{slot}}} N_{\mathrm{sucbit},m}(n_{\mathrm{slot}})}{N_{\mathrm{slot}} \cdot T_{\mathrm{slot}}} \:\:\: [\mathrm{Mbps}].
\end{eqnarray}

Fig.~\ref{fig:sys_throughput} shows the \ac{CDF} of system throughput for average received \ac{SNR} values of $4$, $12$, and $20$ dB.
The following methods for predicting the post-\ac{MUD} \ac{MI} are compared, where (*) specifies the detector employed as the \ac{MUD} (either \ac{LMMSE} or \ac{EP}):
\begin{itemize}
    \item ZF-based SINR (*): This scheme assumes equalization based on the \ac{LS} criterion and predicts the \ac{MI} using the effective \ac{SINR} of each subcarrier~\cite{Shikida2016,Doi2023}. 
    It is a well-known prediction approach employed in the classical \ac{MCS} selection framework and serves as a reference to evaluate the effectiveness of the proposed scheme. This approach is equivalent to using a \ac{ZF} detector as the \ac{MUD}, in which case the post-\ac{MUD} \ac{SINR} is analytically computed in closed form from the \ac{CSI} and the average received \ac{SNR} as
    \begin{equation}
        \gamma_{m, \mathrm{ZF}}[\ell] = \rho \cdot \frac{1}{\left[\left(\bm{H}^{\mathsf{H}}[\ell] \bm{H}[\ell]\right)^{-1}\right]_{m,m}}.
        \label{eq:SINR_ZF}
    \end{equation}
    For the \ac{TB}-level channel quality metric, the average effective \ac{SINR} is used. 
    This is obtained by converting the computed per-subcarrier effective \ac{SINR} values into transmission rates based on Shannon capacity, averaging them over all subcarriers, and then converting the result back into effective \ac{SINR}:
    \begin{subequations}
    \label{eq:ave_SINR}
    \begin{eqnarray}
    \bar{\gamma}_{m,\mathrm{ZF}}
    \!\!&=&\!\!
    \phi^{-1}\left(\bar{R}_m \right), \\
    \bar{R}_m
    \!\!&=&\!\!
    \frac{1}{L} \sum_{\ell = 1}^{L} \phi\left(\gamma_{m,\mathrm{ZF}}[\ell] \right),
    \end{eqnarray}
    \end{subequations}
    where $\phi(x)\triangleq \log_2 \left( 1 + x \right)$.
    \ac{MCS} selection is then performed by mapping the average \ac{SINR} in \eqref{eq:ave_SINR} to \ac{BLER} using the \ac{SNR}–\ac{BLER} characteristics of an \ac{AWGN} channel, and selecting, for each \ac{UE}, the highest \ac{MCS} index that satisfies the reference \ac{BLER}~\cite{Shikida2016,Doi2023}.
    \item LMMSE-based MI (*): This scheme employs the \ac{MI}-based \ac{MCS} selection framework described in Section III, with the workflow diagram shown in Fig.~\ref{fig: MI-based MCS selection analytical}. The \ac{MI} prediction method in \eqref{eq:PreMI_LMMSE} is adopted~\cite{Jensen2010}, where the prediction accuracy of the post-\ac{MUD} \ac{MI} corresponds to Fig.~\ref{fig:res_Mean}, and its mapping to post-decoding \ac{BLER} corresponds to Fig.~\ref{fig: MI-BLER}.
    \item VSS-based MI (*): This scheme incorporates the \ac{VSS}-based \ac{MI} prediction described in Section \ref{sec: Proposed Method}, with the workflow diagram shown in Fig.~\ref{fig: MI-based MCS selection proposed}. Post-\ac{MUD} \ac{MI} prediction is performed using the \ac{VDB} and \ac{ANN}, where the prediction accuracy of the post-\ac{MUD} \ac{MI} corresponds to Fig.~\ref{fig:res_VDB}, and its mapping to post-decoding \ac{BLER} corresponds to Fig.~\ref{fig: MI-BLER} (d)-(f).
\end{itemize}
%
%

We examine the results in Fig. \ref{fig:sys_throughput}.
First, for the \ac{ZF}-based \ac{SINR} scheme~\cite{Doi2023}, the system throughput markedly degrades, especially in the low-\ac{SNR} regime.
There are two causes: (i) reduced accuracy of post-\ac{MUD} \ac{MI} prediction due to the detection-performance gap between the \ac{LMMSE} and \ac{ZF} detectors, and (ii) \ac{BLER}-mapping error introduced by the averaging operation via Shannon capacity, which is not aligned with transmission using discrete constellations.
Consequently, although this framework is widely adopted regardless of the underlying \ac{MUD}, it often fails to assign an appropriate \ac{MCS} in practice.
Many practical systems gradually mitigate this issue using correction mechanisms such as \ac{OLLA}~\cite{Durán2015,Peralta2022,Doi2023,Liu2023}; however, the throughput during the initial phase of link adaptation tends to be significantly reduced.
These results also suggest that when \ac{OLLA} underperforms, or in mission-critical communications, the conventional approach may fail to ensure highly reliable communication.
Next, with the \ac{LMMSE}-based \ac{MI} scheme (LMMSE), substantial throughput gains are observed over the conventional \ac{ZF}-based method, despite using the same \ac{MUD}.
This improvement stems from analytical post-\ac{MUD} \ac{MI} prediction and accurate mapping to post-decoding \ac{BLER} via \ac{EXIT} curves, which enable proper \ac{MCS} selection.
Specifically, the average throughput improves by approximately (a) $40$ Mbps at $4$ dB, (b) $20$ Mbps at $12$ dB, and (c) $5$ Mbps at $20$ dB.
The smaller gain at high \ac{SNR} arises because the performance gap between the \ac{LMMSE} and \ac{ZF} detectors narrows in this region.

The most notable observation is that changing the \ac{MUD} from \ac{LMMSE} to \ac{EP} does not improve the system throughput when \ac{MCS} selection remains based on the \ac{LMMSE} criterion (\textit{i.e.}, the \ac{LMMSE}-based \ac{MI} scheme with \ac{EP}).
This indicates that merely enhancing the detector does not translate into throughput gains unless the \ac{MCS} selection strategy is co-designed with the adopted \ac{MUD}.

Finally, when the \ac{EP} detector is used as the \ac{MUD} and the proposed \ac{VSS}-based \ac{MI} prediction is incorporated into \ac{MCS} selection, the system throughput improves significantly across all \acp{SNR}.
Relative to the \ac{LMMSE}-based \ac{MCS} assignment, the average improvements are approximately (a) $10$ Mbps at $4$ dB, (b) $15$ Mbps at $12$ dB, and (c) $25$ Mbps at $20$ dB.
These results demonstrate that the proposed framework is crucial for leveraging advanced iterative \ac{MUD} in practical systems.
%

\begin{figure}[!t]
\begin{center}
    \subfloat[\ac{SNR} $=4$ dB.]{
    \includegraphics[width=1.0\columnwidth,keepaspectratio=true]{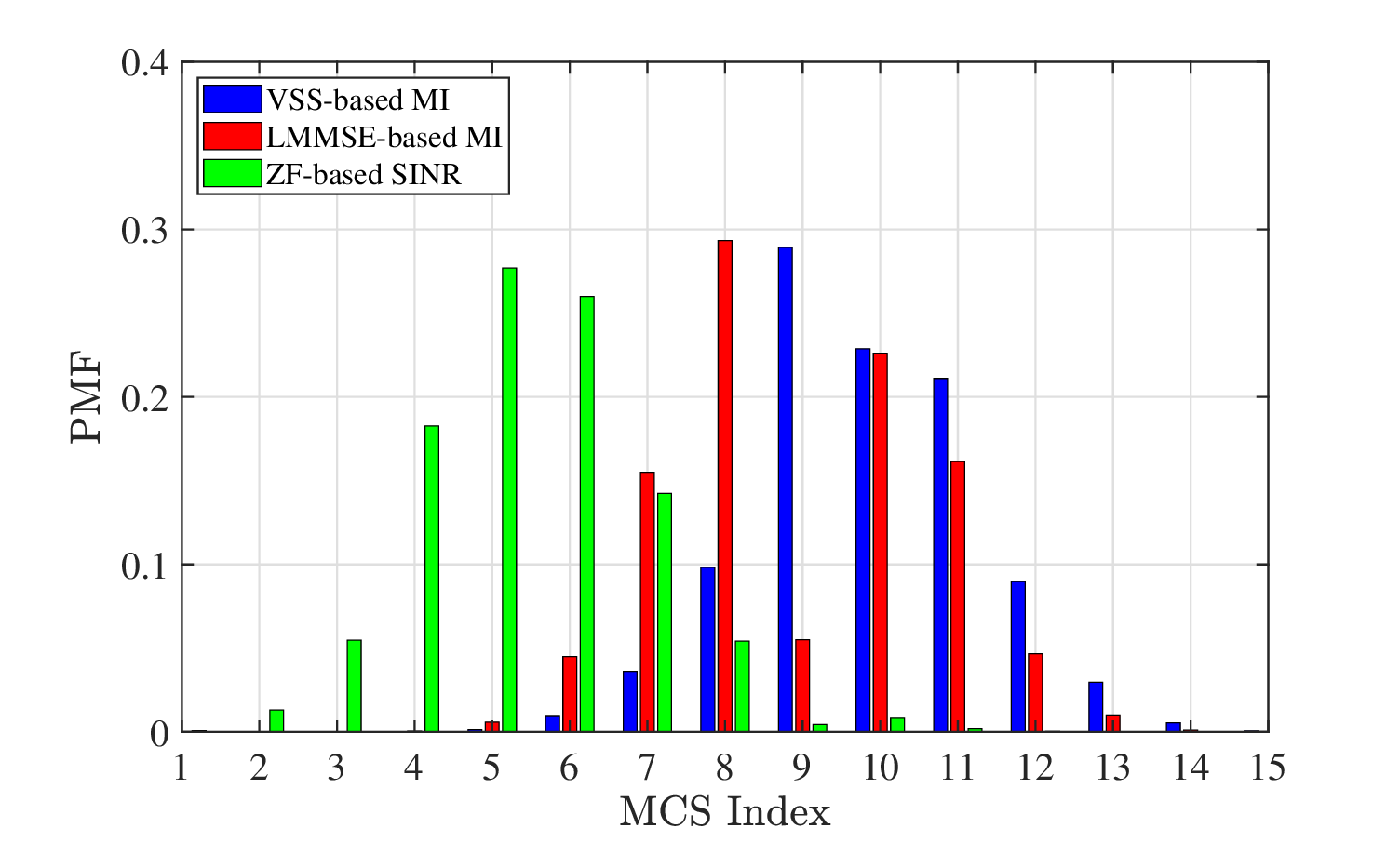}
    }
    \\
    \vspace{-4mm}
    \subfloat[\ac{SNR} $=12$ dB.]{
    \includegraphics[width=1.0\columnwidth,keepaspectratio=true]{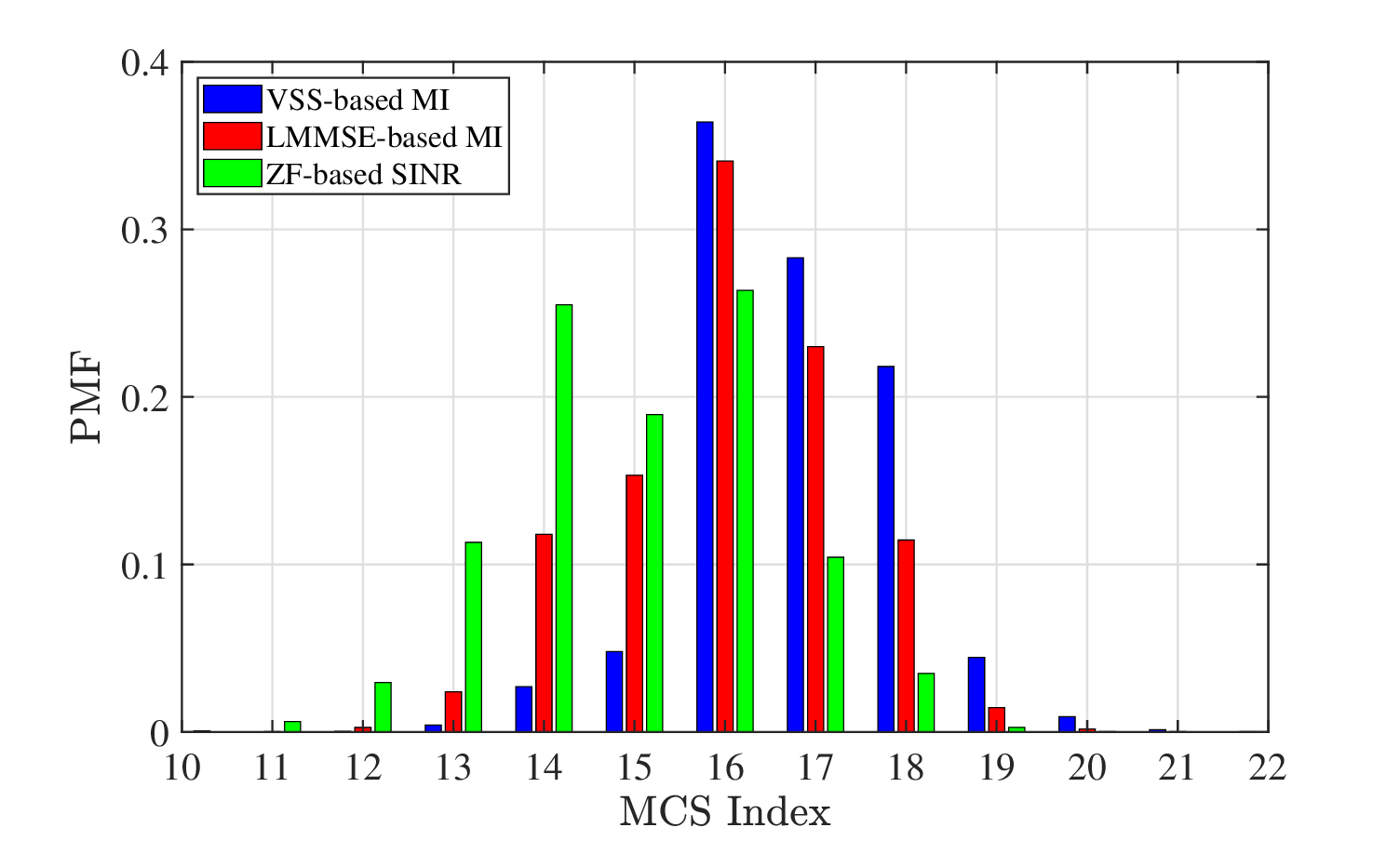}
    }
    \\
    \vspace{-4mm}
    \subfloat[\ac{SNR} $=20$ dB.]{
    \includegraphics[width=1.0\columnwidth,keepaspectratio=true]{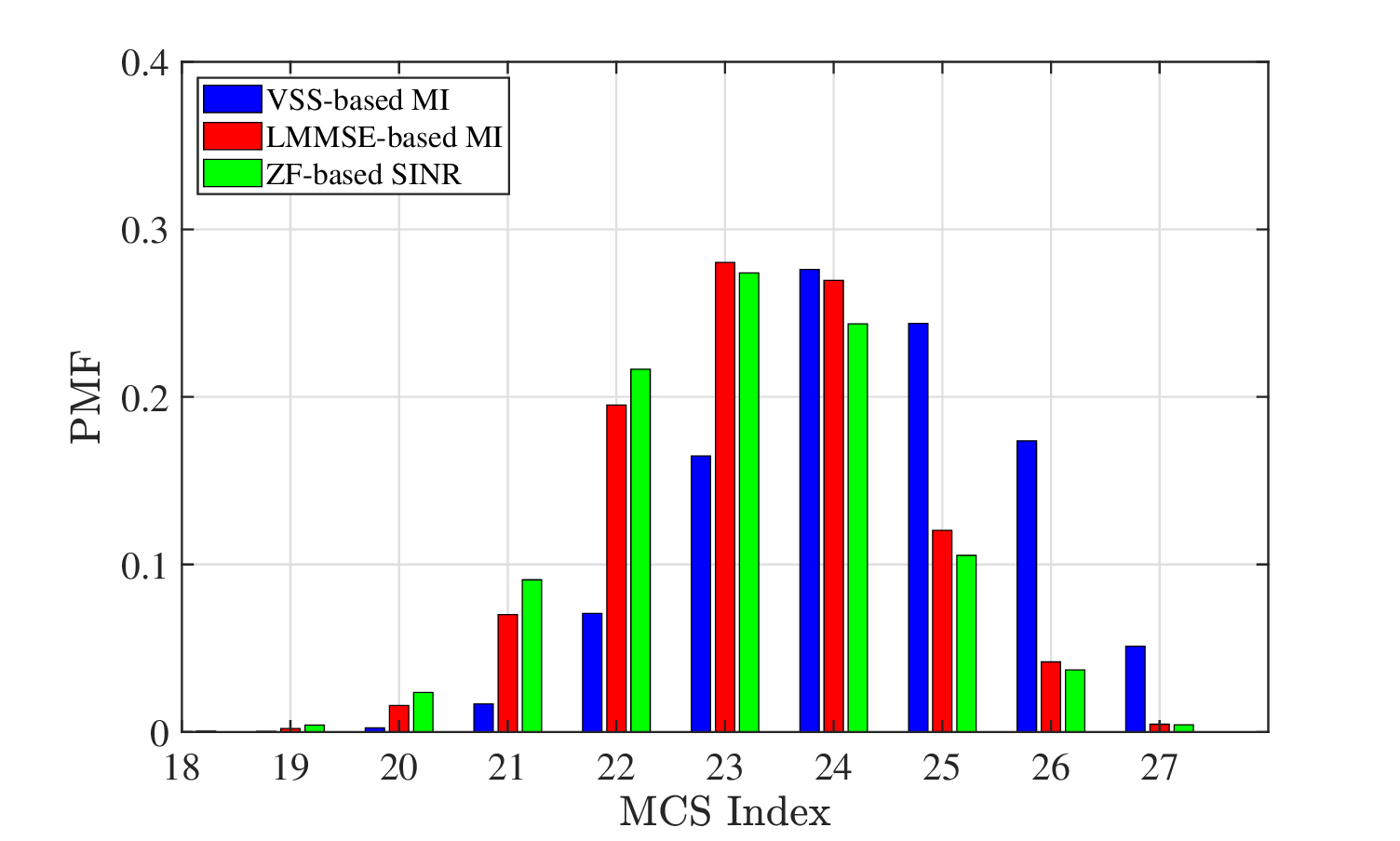}
    }
    \caption{\acs{PMF} of \ac{MCS} indices assigned to \acp{UE}.}
    \label{fig:PMF_MCS}
\end{center}
\vspace{-6mm}
\end{figure}

Fig. \ref{fig:user_throughput} shows the \ac{CDF} of user throughput for the same simulation setup as in Fig.~\ref{fig:sys_throughput}.
This allows us to observe variations in throughput for individual \acp{UE} under different channel conditions, which cannot be captured when only the aggregated system throughput is considered.
An important insight from these results is that the proposed scheme achieves throughput improvements across all \ac{SNR} and rate regions.
In other words, the proposed framework does not improve the rates of certain \acp{UE} at the expense of degrading others but rather provides gains for all \acp{UE}.
This property is highly desirable in terms of \ac{UE}
fairness in more realistic system-level simulations.
The overall trend is consistent with the system throughput results, indicating that the proposed scheme also offers substantial performance improvements over the conventional \ac{ZF}-based method from the perspective of user throughput.
Moreover, the gain from using the \ac{VSS}-based \ac{MI}
prediction is most pronounced in the high \ac{SNR} region, where the iterative gain of the \ac{EP} detector is significant.

Fig. \ref{fig:PMF_MCS} shows the \ac{PMF} of the \ac{MCS} indices assigned to each \ac{UE} in the above simulations.
With \ac{VSS}-based \ac{MI} prediction, the \ac{MCS} selection reflects the superior detection performance of \ac{EP}-based \ac{MUD}, which is especially evident for higher-order modulations where the performance gap between the \ac{LMMSE} and \ac{EP} detectors is large.
This observation explains the throughput improvements seen in the previous results.
Notably, the change in the \ac{PMF} resulting from different post-\ac{MUD} \ac{MI} prediction methods is not merely a simple parallel shift.
%
This fact suggests that the proposed method improves the accuracy of the predicted \ac{MI} at a fundamentally different level, rather than merely applying a fixed backoff correction to the predicted values.


\section{Discussion and Future Work}
\label{sec: FutherDiscussion}

In Section~\ref{sec: Throughput Evaluation}, simulation results were presented under the assumption of a long-term statistically static environment after resource allocation.
%
To prepare for the practical use and further development of the proposed method, it is important to comprehensively evaluate its performance in more realistic system-level simulations that incorporate long-term variations in channel statistics, integration with \ac{OLLA}, and scheduling mechanisms responsible for resource allocation.
Prior to such system-level evaluations, this section provides supplementary discussion and insights into the feasibility of the proposed method in practical systems and the potential challenges associated with integrating these components.

\subsection{Long-Term Variations in Channel Statistics}


In practical operation, the \ac{VDB} deployed at a \ac{BS} is expected to be constructed based on a combination of simulation data and real measurement data collected during a warm-up phase, thereby tailoring it to the surrounding propagation environment.
As a result, the occurrence of channels that significantly deviate from those stored in the \ac{VDB} is unlikely to be both random and frequent.
This tendency is particularly pronounced in low-frequency bands such as sub-6 GHz, where long-term environmental variations that affect channel models generally evolve slowly.
Under such conditions, the proposed framework is expected to maintain high prediction accuracy stably over extended periods of operation.
This expectation is further supported by the results in Subsection~\ref{subsec:VSS_MI_res}, where high prediction accuracy was achieved even though the evaluation employed channel realizations that were generated independently of those used for \ac{VDB} construction and included medium- and short-term channel fluctuations.

Nevertheless, it is also possible that long-term channel statistics may change significantly due to gradual or abrupt environmental variations.
Even in such scenarios, the proposed method is expected to maintain relatively high prediction accuracy or to recover its accuracy within a short period through \ac{VDB} updates.
This is because, as shown in Fig.~\ref{fig: Accuracy (VDB size)}, the proposed approach can achieve an accurate prediction with a relatively small \ac{VDB}, implying that the gradual addition and updating of \ac{VDB} entries enable flexible adaptation to environmental changes.
Immediately after a change in channel statistics, the system performs a prediction based on the most similar entries stored in the \ac{VDB}, allowing rapid adaptation even to previously unseen environments.
This capability to realize such flexibility through localized database updates constitutes one of the key advantages of the \ac{VDB}-based framework and represents a significant benefit over conventional learning-based approaches, which typically rely on retraining.

\subsection{Integration with OLLA}

    

The proposed framework can be seamlessly integrated into system-level simulations as an \ac{ILLA} mechanism and operated in conjunction with conventional \ac{OLLA}.
In this configuration, it improves the prediction accuracy of the demodulator output \ac{MI} at periodic \ac{CSI} acquisition instances and enables appropriate initial \ac{MCS} selection for each \ac{UE}.
As a result, block error occurrences—particularly in the initial stage—can be suppressed, thereby reducing the retransmission probability and contributing not only to throughput improvement but also to more stable \ac{OLLA} convergence.
A particularly relevant application scenario for the proposed method is \ac{URLLC}.
In general, since target \ac{BLER} values are set to be very low in high-reliability communications, the update step size of \ac{OLLA} becomes extremely small, which may cause \ac{OLLA} alone to fail to track channel variations adequately or even to converge~\cite{Peralta2022}.
Moreover, under stringent latency constraints, operating \ac{OLLA} alone—while implicitly assuming the occurrence of transmission errors—is not necessarily appropriate.
Under such conditions, integrating the proposed framework with \ac{OLLA} as an \ac{ILLA} mechanism can improve the accuracy of initial \ac{MCS} selection, thereby enabling both enhanced system performance and improved stability.

On the other hand, to ensure optimal operation of this integrated scheme, careful design of \ac{OLLA}-related parameters, such as the update step size and control policies for spatial multiplexing, is required.
These issues constitute important directions for future research.

\subsection{Future Challenges and Practical Feasibility}

In system-level simulations, it is necessary to account for the behavior of schedulers responsible for resource allocation across the time, frequency, and spatial domains.
In such scenarios, the resources allocated to each \ac{UE} dynamically vary on a per-slot basis, leading to increased diversity in the channel structures experienced by individual \acp{UE}.
In particular, the combinations of spatially multiplexed \acp{UE} determined by spatial resource allocation have a significant impact on detector performance and, consequently, on communication quality assessment for rate control.

Under such scheduler-coupled environments, the design and operation of a \ac{VDB} tailored to the scheduler behavior become important issues for further investigation.
Specifically, this introduces new challenges, including further refinement of feature vector design, advanced \ac{VDB} construction strategies (\textit{e.g.}, database scale, partitioning methods, and granularity design), and more sophisticated \ac{VSS} processes (\textit{e.g.}, hierarchical search and parameter optimization)~\cite{Hashimoto2025}.
Nevertheless, these challenges do not constitute fundamental technical barriers that undermine the practical feasibility of the proposed framework and are expected to be adequately addressed through systematic future studies.

Overall, these considerations highlight that the proposed framework provides a flexible and extensible foundation for system-level operation in scheduler-driven wireless networks.

\section{Conclusion}
\label{sec: Conclusion}

In this paper, we proposed an \ac{MI}-based \ac{MCS} selection framework that incorporates a \ac{VSS}-based \ac{MI} prediction scheme for massive \ac{MU-MIMO-OFDM} systems employing advanced \ac{MUD}.
The framework designs all \ac{MCS} selection processing based on \ac{TB}-level \ac{MI}, where the mapping from high-accuracy post-\ac{MUD} \ac{MI} to post-decoding \ac{BLER} is enabled through a prediction function tailored to the \ac{MCS} table using \ac{EXIT} curves.
When an \ac{LMMSE} detector, for which the post-\ac{MUD} \ac{MI} can be analytically predicted, is employed, the framework achieves optimal \ac{MCS} selection with respect to the reference \ac{BLER}.
Furthermore, when an \ac{EP} detector is used as an advanced iterative \ac{MUD}, the framework incorporates high-accuracy \ac{MI} prediction based on \ac{VSS}, utilizing \ac{VDB} and \ac{ANN} search, thereby reflecting iterative detection gains in \ac{MCS} selection.
The key advantage of the proposed framework is its applicability to any \ac{MUD} scheme whose decoder input cannot be analytically derived. 
Computer simulation results verify the effectiveness of the proposed method from both system and user throughput perspectives in a setup compliant with the \ac{5G NR} standard.
The results demonstrate that the proposed \ac{MCS} selection framework can effectively translate the performance gains of advanced iterative detectors into tangible system-level improvements and that it can significantly enhance reliability and spectral efficiency.
\bibliographystyle{IEEEtran}
\bibliography{listofpublications}

\end{document}